%% file: main.tex
\documentclass[superscriptaddress,nobibnotes,amsmath,amssymb,notitlepage,twocolumn,pra,longbibliography]{revtex4-2}

\usepackage{bm,braket}
\usepackage[toc,page]{appendix}
\usepackage{comment}
\usepackage{dcolumn}

\usepackage{amsfonts}

\usepackage{graphicx,color,hyperref}
\usepackage[caption=false]{subfig}
\hypersetup{colorlinks=true, linkcolor=blue, citecolor=blue, urlcolor=blue} 

\graphicspath{{./img/}}

\newcommand{\beq}[1]{\begin{equation}\label{#1}}
\newcommand{\eep}{\;.\end{equation}}
\newcommand{\eec}{\;,\end{equation}}
\newcommand{\eeq}{\end{equation}}

\newcommand{\om}{\omega}

\DeclareMathAlphabet{\mathcal}{OMS}{cmsy}{m}{n} % Changes font for mathcal but leaves the rest of the math fonts in Times.

\renewcommand{\vec}[1]{{\bf #1}}

\usepackage{amsmath}
\usepackage{amssymb}
\usepackage{xcolor}
\usepackage{bbm}
\usepackage{physics}
\usepackage{float}
\usepackage{graphicx}
\usepackage{dcolumn} % Align table columns on decimal point
\usepackage{bm} % bold math
\usepackage{siunitx}
\usepackage{enumitem}  % to use (i) etc for enumerate lists
\usepackage{pifont}
\usepackage{mathrsfs}

\makeatletter
\renewcommand*{\fnum@figure}{{\normalfont\bfseries \figurename~\thefigure}}
\makeatother

\allowdisplaybreaks

\definecolor{orange}{rgb}{1,0.5,0}

\newcommand{\sect}[1]{\vspace{0.3em}{\it #1.}---}

\graphicspath{{./img/}}

\DeclareMathAlphabet{\mathcal}{OMS}{cmsy}{m}{n} % Changes font for mathcal but leaves the rest of the math fonts in Times.

\newcommand{\intBZ}{\int_{\text{BZ}}} % Integration over BZ

\makeatletter
\newcommand{\specificthanks}[1]{\@fnsymbol{#1}}% Inserts a specific \thanks symbol
\makeatother

\begin{document}

\preprint{APS/12Three-QED}

\title{Topological Acoustic Diode}

%%%% AFFILIATIONS %%%
\newcommand{\TCM}{{Theory of Condensed Matter Group, Cavendish Laboratory, University of Cambridge, J.\,J.\,Thomson Avenue, Cambridge CB3 0HE, UK}}
\newcommand{\UoM}{Department of Physics and Astronomy, University of Manchester, Oxford Road, Manchester M13 9PL, UK}
%%% AUTHORS %%%

%%% AUTHORS %%%

\author{Ashwat Jain}
\email{ashwat.jain@postgrad.manchester.ac.uk}
\affiliation{\UoM}

\author{Wojciech J. Jankowski}
\email{wjj25@cam.ac.uk}
\affiliation{\TCM}

\author{M. Mehraeen}
\email{mandela.mehraeen@manchester.ac.uk}
\affiliation{\UoM}

\author{Robert-Jan Slager}
\email{robert-jan.slager@manchester.ac.uk}
\affiliation{\UoM}
\affiliation{\TCM}
\date{\today}

\begin{abstract}
We show that certain three-dimensional topological phases can act as acoustic diodes realizing nonlinear odd acoustoelastic effects. Beyond uncovering topologically-induced anomalous acoustic second-harmonic generation and rectification, we demonstrate how such nonlinear responses are uniquely captured by the momentum-space nonmetricity tensor in the quantum state Hilbert-space geometry. In addition to completing the classification of quantum geometric observables in the quadratic response regime, our findings reveal unexplored avenues for experimental realizations of acoustic diodes using effective $\theta$~vacua of axion insulators adaptable for topological engineering applications.  
\end{abstract}

\maketitle

%Nonlinear responses and diodes (and acoustic)

\sect{Introduction}Nonlinear phenomena constitute a~growing field of interest, underpinned by various theoretical and experimental advances~\cite{Konotop2016, ortix2021nonlinear, ideue2021symmetry, du2021nonlinear, nagaosa2024nonreciprocal, shim2025spin, suarez2025nonlinear, Jiang2025}. Fundamentally, these effects arise from anharmonicities in various forms of matter subject to time-dependent forcing, which can notably culminate in second-harmonic generation (SHG), manifested by a frequency doubling in responses~\cite{Boyd2019}. Among such nonlinear effects, of particular technological interest are rectifications, that is conversions of ac currents to dc currents, which are definitional to the operating principle of diodes. Although such phenomena in matter typically arise as a response to electromagnetic stimulation, similar nonlinear responses can also arise upon perturbing crystals with acoustic waves, or sound~\cite{liang2009acoustic, liang2010acoustic}. Recent advances within this scope involve acoustoelectric effects, such as quantum nonlinear acoustic Hall effects~\cite{Su2025}. 

%Topologies for nonlinearities and axions

Experimentally accessible and fundamentally intriguing topological materials~\cite {Qi2008, Hasan2010rmp, moore2010birth, Qi2011rmp, Kruthoff2017,Po2017,Bradlyn2017, Weylrmp, tokura2019magnetic, bernevig2022progress} are particularly promising for realizing unconventional nonlinear effects. Exotic nonlinearities in responses of topological matter can be captured by current correlators, where intrinsic effects arise at the level of one-loop diagrams via propagators of topological quasiparticles and their coupling to external forcing fields, represented by vertices~\cite{Parker2019, Michishita2021effects, Michishita2022dissipation, Mehraeen2025}. In that context, electromagnetic nonlinearities, including optical responses and magnetoresistances, are naturally supported by three-dimensional topological insulators~\cite{Qi2008, Sodemann2015, yasuda2016large, he2021quantum, mehraeen2024proximity}, such as axion insulators~\cite{Essin2009, Sekine2021}, and topological semimetals such as Weyl and Dirac materials~\cite{Weylrmp, morimoto2016chiral, shvetsov2019nonlinear, dzsaber2021giant, kumar2021room, zhang2022large}. In particular, nonlinear optical responses manifested as rectified currents culminate in topological bulk photovoltaic effects~\cite{Alexandradinata2024}, such as quantized circular photogalvanic effects due to Berry curvature~\cite{deJuan2017}, or shift responses due to Hermitian connections~\cite{Ahn2020, Ahn2021} and band torsion~\cite{Jankowski2024PRL}. Formally, exotic nonlinear responses may be phrased in terms of broader classes of quantum geometric tensors (QGTs)~\cite{Ahn2020, Ahn2021, bhalla2022resonant, Bouhon2023, Jiang2025, Mehraeen2025}, demonstrating a deeper interplay of momentum-space Riemannian geometry and quantum response theory~\cite{Ahn2021}. While the original formulations of quantum geometric quantities date back to the last century~\cite{Provost1980}, their physical manifestations beyond correlated~\cite{Peotta2015} or optical phenomena~\cite{Ahn2021, Jankowski2025optical} have initiated several current research directions~\cite{Torma2023, Jiang2025}.

\begin{figure}[t!]
    \centering
    \def\svgwidth{\linewidth}
    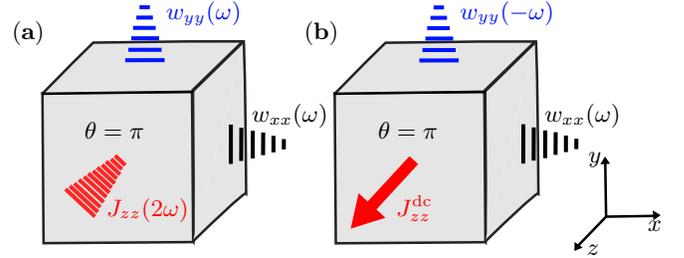
    \caption{Topological acoustic diode responses. {\bf (a)} Odd acoustic second-harmonic generation. {\bf (b)} Odd acoustic rectification. The nonlinear acoustic responses to time-dependent distortion fields $w_{ij}(\omega)$ are topologically-induced by the $\theta$-vacuum of an axion insulator ($\theta = \pi$). The acoustoelastic frequency-doubling $J_{zz}(2\omega)$ and rectified $J^\text{dc}_{zz}$ currents flow in the $z$ direction, orthogonally to the mutually perpendicular forcing directions $(x,y)$.}
    \label{fig:schematics}
\end{figure}

%Quantum geo

Moving beyond the domain of electromagnetic responses, QGTs have recently been shown to manifest in nonlinear viscoelastic phenomena~\cite{Jain2025NOVE}. Notably, such nonlinear elastic responses involve unique geometric structures in the parameter spaces of deformation fields, rather than single-particle momenta~\cite{Avron1995, Klevtsov2015, Jain2025NOVE}. Such nontrivial quantum geometries can be induced by topological invariants, although the according viscoelastic effects go beyond the strict presence of topology. In particular, these effects persist upon breaking the symmetries protecting the nontrivial topologies. Elastic nonlinearities can therefore galvanize uncharted avenues for further exploration of quantum responses that are driven by unprecedented gauge-invariant, and thus measurable, geometric tensor quantities.

%In this work

In this work, motivated by the rich quantum geometry of topological materials, we identify a set of odd nonlinear acoustoelastic effects adaptable for constructions of topological acoustic diodes. As a concrete realization of such a topological acoustic diode, we propose acoustically driven axion insulators, phases of matter that have been experimentally retrieved recently~\cite{Otrokov2019, Qiu2023, Qiu2025}. We demonstrate that the $\theta$~vacua of axion insulators not only realize odd acoustic rectifications, central to the operation of acoustic diodes, but also demonstrate odd acoustic SHG. We fully characterize the quantum-geometric nature of this acoustic SHG response of topological electrons driven by the time-dependent external acoustic deformations.

More fundamentally, we uncover the central role of the quantum nonmetricity tensor, reflecting the quantum metric incompatibility of the Hermitian connection over the manifold of cell-periodic Bloch states. This provides a capstone to the set of Riemannian-geometric tensors manifested in quadratic material responses. Beyond retrieving uncharted acoustoelastic phenomenology, we showcase how the odd responses of an axionic topological acoustic diode can be experimentally probed and adapted for topological engineering purposes. Finally, we discuss how the realization of identified odd acoustoelastic effects may impel further pursuits in terms of broader applications of topological materials.

\sect{Results}In the following, we introduce and characterize the nonlinear odd acoustoelastic effects in a topological acoustic diode. We begin by considering frequency components of distortion fields ${w_{ij}(\omega) = \frac{\partial u_{i}(\om)}{\partial x_j}}$, with elastic displacements ${x_i \rightarrow x_i + u_{i}(\om)}$ in spatial coordinates~$x_i$, and frequencies $\omega$ under time-dependent deformations $u_i(t)= \int \text{d}\omega~e^{\text{i} \omega t} u_i(\omega)$. Distortions are then implemented by gauging momenta using a generalized Peierls substitution $k_j \rightarrow k_j - w_{ij}\sin(a k_i)/a$ with $a$ the lattice constant~\cite{Shapourian2015,Jain2025NOVE}. The nonlinear acoustoelastic currents in response to time-dependent distortion fields $w_{kl}(\omega_1)$ and $w_{mn}(\omega_2)$ read
\begin{equation}
    J_{ij}(\Omega)=\eta^{}_{ij;kl,mn}(\omega_1,\omega_2) w_{kl}(\omega_1) w_{mn}(\omega_2),
\end{equation}
with the nonlinear acoustic response function given by $\eta^{}_{ij;kl,mn}(\omega_1,\omega_2)$ and Einstein summation implied over spatial coordinate indices. This results in elastic momentum currents $J_{ij}(\Omega)$ with frequency $\Omega =\omega_1 + \omega_2$, for any pair of input frequencies $\om_1, \om_2$. In particular, we focus on odd acoustic SHG ($\omega_1 = \omega_2$), and dc rectified responses $(\Omega=0)$.

In gapped phases, the corresponding nonlinear odd SHG at sub-gap frequencies, and rectification response functions at arbitrary frequencies, which formally includes contributions from all one-loop diagrams [see the Supplemental Material (SM)~\cite{SI}], are of the  form
\begin{align}
    \eta^{\mathrm{odd,\,SHG}}_{ij;kl,mn}(\omega,\omega) &=\int_{\textbf{k}} \Big[\frac{1}{\omega} \alpha_1 + \alpha_2 + \omega \alpha_3 \Big] \label{eq:eta_SHG} ,\\
    \eta_{ij;kl,mn}^{\mathrm{odd,\,rect}}(\omega,-\omega) &= \int_{\textbf{k}} \bigg[\delta(\Delta^{ba}-\omega)\alpha_e  + \delta(\Delta^{ba}+\omega)\alpha_h\bigg]\label{eq:eta_rect},
\end{align}
with $\int_{\textbf{k}} \equiv \intBZ \frac{\text{d}^3 \textbf{k}}{(2\pi)^3}$ a three-dimensional integral over the Brillouin zone (BZ). We present explicit expressions for $\alpha_i$ in Table~\ref{tab:alphas}. The nonlinear response to dc fields~\cite{Jain2025NOVE} can be retrieved in the zero-frequency limit of the SHG response, reading $\eta^{\mathrm{odd,\,dc}}(0,0) = {\int_\textbf{k}~[-\text{i}\tau \alpha_1 + \alpha_2]} \label{eq:eta_dc}$, with $\tau$ the characteristic scattering time of bulk electrons. In Fig.~\ref{fig:schematics}, we showcase the nonlinear phenomenologies of introduced odd acoustic SHG and rectified responses in the specific case of an axion insulator model~\cite{Wieder2018}, whose specific formulation is detailed in the End Matter.

\begin{figure}
    \centering
\def\svgwidth{\linewidth}
    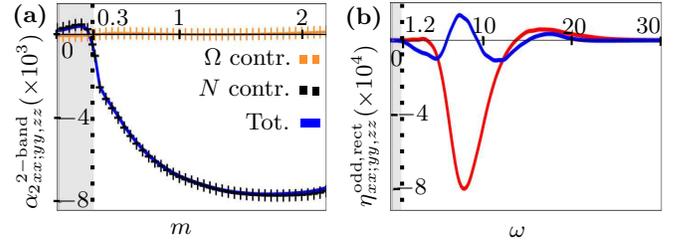
    \caption{Nonlinear acoustic responses of an axion insulator diode. {\bf (a)} Odd second-harmonic response term $\alpha^{2-\text{band}}_{2,xx;yy,zz}$ against the spin-splitting mass term ($m$) controlling the band gap. {\bf (b)} Odd rectification ($\eta^{\text{odd,rect}}_{xx;yy,zz}$) response of a topological acoustic diode vs. frequencies $\omega$. The real (\textit{blue}) and imaginary (\textit{red}) parts of the nonlinear acoustic responses reflect distinct dependences on underlying quantum geometries. In the acoustic second-harmonic susceptibility, the nonmetricity tensor ($N$) contribution dominates the Berry curvature ($\Omega$) term within the total response (\text{Tot.}).}
    \label{fig:fig2}
\end{figure}

\begin{table*}[t]
\centering
\begin{ruledtabular}
\begin{tabular}{c c c}
& \textbf{Two-band contribution} & \textbf{Three-band contribution} \\
\hline\\[-1ex]
$\alpha_1$& $\frac{4\text{i}}{3}\sum\limits_{a,b}f_{ba}\bigg[g_{[(kl,mn)}^{ba}\partial_{ij]}\Delta^{ba} \bigg]$
& \ding{55}\\[3ex]
$\alpha_2$& $\sum\limits_{a,b}f_{ba}\bigg[\frac{\Omega_{ij,(kl}^{ba}\partial_{mn)}\Delta^{ba}}{\Delta^{ba}} + \frac{32\text{i}}{3}N_{[(kl,mn),ij]}^{ba}\bigg]$
& $\frac{2}{3} \sum\limits_{a,b,c} \frac{\text{Re}Q_{ij,(kl,mn)}^{abc}}{\Delta^{ba}}\bigg(2+\frac{(\Delta^{ba})^2}{\Delta^{bc}\Delta^{ca}}\bigg)(f_a \Delta^{bc}+f_b \Delta^{ca}+f_c \Delta^{ab})$ \\[3ex]
$\alpha_3$&  $\frac{16\text{i}}{3}  \sum\limits_{a,b}\frac{f_{ba}}{(\Delta^{ba})^2}\bigg[g_{[(kl,mn)}^{ba}\partial_{ij]}\Delta^{ba}\bigg]$
& $
{\scriptsize\begin{aligned} \frac{10\text{i}}{3} \sum\limits_{a,b,c}&\frac{\text{Im} Q_{ij,(kl,mn)}^{abc}}{(\Delta^{ba})^2}\bigg( \frac{f_{bc}\Delta^{ca}}{(\Delta^{bc})^2}[2(\Delta^{bc})^2 - \Delta^{bc}\Delta^{ca} -(\Delta^{ca})^2] \nonumber \\[-2ex]&\qquad- \frac{f_{ca}\Delta^{bc}}{(\Delta^{ca})^2}[2(\Delta^{ca})^2 - \Delta^{ca}\Delta^{bc} - (\Delta^{bc})^2]\bigg) \bigg]\end{aligned}}$ \\[5ex]
$\alpha_e$&  $-\pi \sum\limits_{a,b} f_{ba} \bigg(\frac{2}{3}\bigg[g_{[(kl,mn)}^{ba}\partial_{ij]}\Delta^{ba}\bigg] + \frac{\text{\text{i}}}{4}\bigg[\Omega^{ba}_{kl,mn}\partial_{ij}\Delta^{ba}\bigg]\bigg)$
& \ding{55} \\[3ex]
$\alpha_h$&  $-\pi \sum\limits_{a,b} f_{ba} \bigg(\frac{2}{3}\bigg[g_{[(kl,mn)}^{ba}\partial_{ij]}\Delta^{ba}\bigg] - \frac{\text{\text{i}}}{4}\bigg[\Omega^{ba}_{kl,mn}\partial_{ij}\Delta^{ba}\bigg] \bigg)$
& \ding{55} \\[3ex]
\end{tabular}
\end{ruledtabular}
\caption{Quantum geometric contributions to anomalous nonlinear second-order responses of a topological acoustic diode. Here, $(\ldots)$ and $[\ldots]$ represent normalized symmetrizations and antisymmetrizations of index pairs. The individual terms are composed of the gauge-invariant geometric tensors and band energy splittings $\Delta^{ab}$, see SM for derivations~\cite{SI}.}
\label{tab:alphas}
\end{table*}

We first note that the introduced acoustoelastic responses can be universally decomposed in terms of two-band and three-band contributions, which respectively involve quantum geometric quantities of pairs and triplets of quantum states, see Table~\ref{tab:alphas}. We find that the odd two-band response in the considered axion insulator model typically dominates over the three-band response contributions: $\alpha_1$~bears no three-band contribution, and $\alpha_2$ realizes a two-band contribution that dominates the three-band contribution. However, both terms are comparable for $\alpha_3$. The form of three-state contributions is also detailed in the SM~\cite{SI}. Notably, the numerical acoustic SHG responses show a distinctive change in behavior around a crossover value, $m = 0.3$, see Fig.~\ref{fig:fig2}(a). Physically, this occurs when the spin-dependent splitting $(m)$ matches the orbital-dependent splitting $(M)$, and the lower two bands become degenerate at the $\Gamma$ point of the BZ. The crossover of the orbital- and spin-splitting mass terms introduces a turning point in the quantum geometric terms that underpin the odd nonlinear acoustoelastic responses. 

As a central result of this work, we also find that the nonlinear odd acoustoelastic SHG response, $\alpha_2$, of the axionic insulator to linearly polarized perturbations is dominated by the nonmetricity ($N$) contribution, with the geometric Berry curvature $(\Omega)$ contribution being negligible, even though the system explicitly breaks time-reversal symmetry ($\mathscr{T}$), while preserving the inversion symmetry ($\mathscr{P}$). The nonmetricity tensor is defined as
\begin{equation}
N_{ij,kl,mn}^{ab}
= -\tilde{\nabla}_{mn} g_{ij,kl}^{ab},
\end{equation}
where $g_{ij,kl}^{ab}$ is the two-state quantum metric tensor in the parameter space of distortions and $\tilde{\nabla}_{mn}$ is the covariant derivative corresponding to the deformation-space Hermitian connection ($\tilde{\Gamma}$) (see SM~\cite{SI}). Geometrically, the nonmetricity tensor measures the failure of the Hermitian connection to parallel-transport the quantum metric along affinely-parametrized geodesics of the Levi-Civita connection, and thus provides a quantitative measure of how distances and angles vary along momentum-space geodesics. Notably, the nonmetricity averaged over momentum space can be directly extracted from the response of any system modeled by a $\mathscr{PT}$-symmetric effective two-band Hamiltonian to linearly polarized perturbations, where contributions of all the other terms listed in Table~\ref{tab:alphas} vanish by definition. 

\sect{Discussion}We now discuss and further examine the structure of the odd responses of an axionic topological acoustic diode. %We then conclude with an outlook for the retrieved odd acoustic rectification responses in a broader context of topological devices and engineering applications.
First, we note that the retrieved responses are highly dependent on the polarization of the perturbing acoustic waves. Under the decomposition of the responses into terms with distinct frequency scaling [Eqs.~\eqref{eq:eta_SHG}--\eqref{eq:eta_rect}], the coefficients $\alpha_1$ and $\alpha_3$ are purely imaginary while $\alpha_2$ is purely real, and thus linearly polarized elastic perturbations selectively probe~$\alpha_2$. Notably, the identified nonlinear odd acoustic responses also depend on band torsion, which is captured by the three-state QGT term (see also SM~\cite{SI}). 

We further compare two distinct identified types of~nonlinear acoustic phenomena. Unlike the acoustic SHG, the rectification response vanishes below the band gap due to the resonant conditions imposed by the delta functions in Eq.~\eqref{eq:eta_rect}. Similarly to dc nonlinear odd viscoelastic effects~\cite{Jain2025NOVE}, the net time-dependent odd responses satisfy sum rules ${\eta_{xx;yy,zz} + \eta_{yy;zz,xx} + \eta_{zz;xx,yy} = 0}$, and similarly $\eta_{xy;yz,zx} + \eta_{yz;zx,xy} + \eta_{zx;xy,yz} = 0$. Under the Peierls gauge substitution, a related correspondence between the shear and normal deformation stresses holds for the dc and SHG responses: $\eta_{xx;yy,zz} = \eta_{zx;xy,yz} = \eta_{yx;xz,zy}$, but not for the rectification response. This stems from a combination of the internal permutation symmetry and that $\omega_1 = \omega_2$ for dc and SHG, but $\omega_1 \neq \omega_2$ for rectification. Thus, the odd acoustic rectification (SHG) responses have four (two) independent components amongst $\eta_{xx;yy,zz}$, $\eta_{yy,zz,xx}$, $\eta_{zz;xx,yy}$, $\eta_{xy;yz,zx}$, $\eta_{yz;zx,xy}$, and $\eta_{zx;xy,yz}$. As a result, an external control of the topology and quantum geometry of a three-dimensional phase allows one to tune up to four individual characteristic response parameters within an operational rectification mode of a topological acoustic diode. 

The four-fold tunability of the odd rectification of a~topological acoustic diode opens up a~multitude of quantum engineered applications. The two-state and three-state quantum geometries can, for instance, be controlled with external magnetic fields, which determine the effective spin- and orbital-splitting gaps ($m$/$M$)~\cite{Sekine2021}. Within the scope of potential applications of the topological acoustic diodes realizing the retrieved odd nonlinear acoustoelastic effects, we envision ultrasound focussing, ultrasonography, switching and logical devices, high-performance acoustic sensors, and noise control~\cite{liang2009acoustic, liang2010acoustic}. Importantly, the acoustoelastic effects may be studied in material candidates for axion insulators realizing field-theoretic $\theta$~vacua, for instance, recently studied topological antiferromagnetic heterostructures based on the chemical compositions of manganese-doped topological insulators, MnBi$_{2n}$Te$_{3n+1}$~\cite{Mong2010, Otrokov2019, Jo2020, Gao2023}.

\sect{Conclusion}In summary, we uncover a set of nonlinear odd acoustic responses in \mbox{topological insulators}, that includes anomalous acoustoelastic second-harmonic generation and rectified currents. Motivated by the recent material identification of axion insulators as natural candidates for such effects, these unconventional responses hold diverse promises for future applications of such systems as topological acoustic diodes in experimental platforms and related engineered devices.

\sect{Acknowledgements} The authors thank Giandomenico Palumbo and Marco Polini for helpful discussions. A.J.~acknowledges funding from the School of Natural Sciences, University of Manchester. W.J.J.~acknowledges funding from the Rod Smallwood Studentship at Trinity College, Cambridge. M.M.~was funded by an EPSRC ERC underwrite Grant No.~EP/X025829/1. R.-J.S. acknowledges funding from an EPSRC ERC underwrite Grant No.~EP/X025829/1, and a Royal Society exchange Grant No. IES/R1/221060, as well as Trinity College, Cambridge. This research was supported in part by grant NSF PHY-2309135 to the Kavli Institute for Theoretical Physics~(KITP).

\section*{End Matter}

\begin{table}
\centering
\begin{tabular}{l r}
\hline \hline \\[-2ex]
Name & Definition \\
\hline\\[-1ex]
Off-diagonal Berry conn.
& $A_{ij}^{ab} = \text{i}\bra{a} \partial_{ij} \ket{b}$ \\[1ex]

Diagonal Berry conn.
& $\mathcal{A}_{ij}^{a} = \text{i}\bra{a} \partial_{ij} \ket{a}$ \\[1ex]

Two-state QGT
& $Q_{ij,kl}^{ab} = A_{ij}^{ab}A_{kl}^{ba}$ \\[1ex]

Three-state QGT
& $Q_{ij,kl,mn}^{abc} = A_{ij}^{ab}A_{kl}^{bc}A_{mn}^{ca}$ \\[1ex]

Quantum metric
& $g_{ij,kl}^{ab} = \text{Re} Q_{ij,kl}^{ab}$ \\[1ex]

Berry curvature
& $\Omega_{ij,kl}^{ab} = -2 \text{Im} Q_{ij,kl}^{ab}$ \\[1ex]

Berry cov. derivative
& $\mathcal{D}_{ij} O^{ab}
= \partial_{ij} O^{ab}
- \text{i} [A_{ij} + \mathcal{A}_{ij},O]^{ab}$ \\[1ex]

Hermitian conn.
& $\tilde{\Gamma}_{ij,kl,mn}^{ba}
= A_{ij}^{ab} (\mathcal{D}_{kl} A_{mn})^{ba}$ \\[1ex]

Christoffel symbol
& $\Gamma^{ab}_{ij,kl,mn}
= \partial_{(kl} g_{mn),ij}^{ba} - \frac{1}{2}\partial_{ij} g_{kl,mn}^{ba}
$ \\[1ex]

Torsion
& $\mathcal{T}_{ij;kl,mn}^{ab} = 2\Tilde{\Gamma}_{ij;[kl,mn]}^{ba}$\\[1ex]

Contorsion
& $K_{ij,kl,mn}^{ab}
= \frac{1}{2}\mathcal{T}_{ij,kl,mn}^{ab}
+ \mathcal{T}_{(kl,mn),ij}^{ab}$ \\[1ex]

Nonmetricity
& $N_{ij,kl,mn}^{ab}
= -\tilde{\nabla}_{mn} g_{ij,kl}^{ab}$ \\[1ex]

Disformation
& $L_{ij,kl,mn}^{ab}
= N_{ij,(kl,mn)}^{ab}
- \frac{1}{2}N_{kl,mn,ij}^{ab}$ \\[1ex]

\hline \hline
\end{tabular}
\caption{Quantum geometric quantities arising in the nonlinear acoustic response functions. The quantum geometric nonmetricity tensor is central to the odd acoustic second-harmonic responses, beyond the deformation-space QGTs~\cite{Avron1995, Klevtsov2015}, Levi-Civita connection~\cite{gao2014field, Hetenyi2023, mehraeen2024quantum, Jain2025PRB}, and torsion tensor~\cite{Ahn2021, Hsu2023, Jankowski2024PRL} defined above.}
\label{tab:geometry}
\end{table}

\sect{Details on the axion insulator models} In the following, we provide further details on the effective models of axion insulators and the odd acoustic SHG and rectifications studied in the main text. The model four-band Hamiltonian for an axion insulator studied for nonlinear acoustic responses in this work reads~\cite{Wieder2018}
\begin{equation}
\label{eq:Hamil}
    \begin{split}
        \mathcal{H}(\vec{k})
        &=~m_0\tau^{z} + m \sigma^{z}
        + M\tau^{z}\sigma^{z}
        \\
        &+
        v_{2,xy}\tau^{x}\sigma^{z}
        [\sin(k_{x}) + \sin(k_{y})] + v_{2,z}\tau^{x}\sigma^{x}\sin(k_{z})
        \\
        &+
        \sum_{i=x,y,z}[t_{1,i}\tau^{z} \cos(k_{i})
        + t_{\text{PH},i} \cos(k_{i})
        \\
        &+ t_{2,i}\tau^{y} \sin(k_{i}) + v_{1,i}\tau^{x}\sigma^{i}\sin(k_{i})],
    \end{split}
\end{equation}
where $\sigma$ and $\tau$ denote spin- and orbital-sector Pauli matrices, Kronecker products are left implicit, and the spin- and orbital-splitting masses are $m$ and $M$, respectively. The model parameters are set to: ${m_0 = -5}$, ${M = 0.3}$, ${t_{\text{PH},x}=t_{\text{PH},y} = 0.3}$, ${t_{\text{PH},z}=0}$, ${t_{x} = 2.3}$, ${t_{y} = 2.5}$, ${t_{z} = 3}$, ${t_{2,x} = 0.9}$, ${t_{2,y} =  t_{2,z} = 0}$,
 ${v_{1,x} = v_{1,y} = 3.2}$, ${v_{1,z} = 2.4}$, ${v_{2,xy} = 1.5}$, ${v_{2,z} = 0.4}$. This model has inversion symmetry ($\mathscr{P}$), while the splitting masses $m$ and $M$ explicitly break the time-reversal symmetry ($\mathscr{T}$), consistently with the conventional symmetries of axion insulators.

In this work, $m$ is varied to study the SHG response, while for the rectified response, we fix $m = 1.2$. The Fermi level is set such that the lower (upper) two bands are fully occupied (unoccupied).

\sect{Details on quantum geometry with nonmetricity}We provide a completed table of quantum geometric quantities on the extended Hilbert space, which capture responses up to quadratic order in perturbations. Beyond the torsion $(\mathcal{T})$ tensor, these involve the momentum-space nonmetricity $(N)$ and disformation $(L)$ tensors. In Table~\ref{tab:geometry}, all relevant geometric entities are summarized, starting from the definitions of the Berry connections in the bundles of Bloch states $\ket{a}$. In particular, the corresponding geometric tensors capture variations of Bloch states over the parameter space of acoustic distortion fields $w_{ij}$, which are generated by variations $\partial_{ij} \equiv \delta/\delta w_{ij}$.

\bibliography{refs}

\end{document}

\title{{\textsc{SUPPLEMENTAL MATERIAL}} \\ Topological Acoustic Diode}

\newcommand{\TCM}{{TCM Group, Cavendish Laboratory, University of Cambridge, J.\,J.\,Thomson Avenue, Cambridge CB3 0HE, UK}}
\newcommand{\UoM}{Department of Physics and Astronomy, University of Manchester, Oxford Road, Manchester M13 9PL, UK}

%%% AUTHORS %%%

\author{Ashwat Jain}
\email{ashwat.jain@postgrad.manchester.ac.uk}
\affiliation{\UoM}

\author{Wojciech J. Jankowski}
\email{wjj25@cam.ac.uk}
\affiliation{\TCM}

\author{M. Mehraeen}
\email{mandela.mehraeen@manchester.ac.uk}
\affiliation{\UoM}

\author{Robert-Jan Slager}
\email{robert-jan.slager@manchester.ac.uk}
\affiliation{\UoM}
\affiliation{\TCM}

\date{\today}

\maketitle

\tableofcontents 
\newpage

\section{Hermitian connection decomposition}
\label{app:geometry}

Here, we briefly review the general decomposition of affine connections on manifolds, see e.g., Refs.~\cite{hehl1995metric, blauGR}, for further details. We then apply this to derive useful properties of the Hermitian connection on the manifold of cell-periodic Bloch states. Consider a general connection $\tilde{\nabla}$, which acts on momentum-space rank-$2$ tensors as
\begin{equation}
    \tilde{\nabla}_{k} \mathcal{O}_{ij}
    =
    \partial_{k} \mathcal{O}_{i j}
    -
    \tilde{\Gamma}_{l i k} \tensor{\mathcal{O}}{^{l}_{j}}
    -
    \tilde{\Gamma}_{l j k} \tensor{\mathcal{O}}{^{l}_{i}}.
\end{equation}
We use the decomposition
$\tilde{\Gamma} = \Gamma_{\text{g}} + C$, where $( \Gamma_{\text{g}} )_{  i j k}
=
\frac{1}{2}
( \partial_{k} g_{i j} 
+
\partial_{j} g_{i k}
-
\partial_{i} g_{j k} )
$
is a component of the Levi-Civita connection $\nabla$, which parallel-transports the metric along momentum-space geodesics,
$\nabla_{  i} g_{j k} = 0$. The metric-incompatibility of $\tilde{\nabla}$ is captured by the nonmetricity tensor~\cite{hehl1995metric, blauGR, Palumbo2024weyl}
\begin{equation}
    N_{i j k}
    \equiv
    - \tilde{\nabla}_{ k} g_{i j}
    =
    2 C_{(i j) k},
\end{equation}
where index (anti-)symmetrizations are normalized as
\begin{subequations}
\label{SM_eq_symmetrization}
    \begin{align}
        \mathcal{O}_{(i_1 \cdots i_n)j}
        &=
        \frac{1}{n!} \sum_{\sigma \in S_n}
        \mathcal{O}_{i_{\sigma(1)} \cdots i_{\sigma(n)}j},
        \\
        \mathcal{O}_{[i_1 \cdots i_n]j}
        &=
        \frac{1}{n!} \sum_{\sigma \in S_n} \text{sgn} (\sigma)
        \mathcal{O}_{i_{\sigma(1)} \cdots i_{\sigma(n)}j}.
    \end{align}
\end{subequations}
$S_n$ denotes the permutation group of $n$ elements and $\sigma$ a permutation of the set $\{1,\cdots,n\}$. In the simplest relevant case ($n=2$), one has
\begin{subequations}
    \begin{align}
        \mathcal{O}_{(ij)}
        &=
        \frac{
        \mathcal{O}_{ij} + \mathcal{O}_{ji}}{2},
        \\
        \mathcal{O}_{[ij]}
        &=
        \frac{
        \mathcal{O}_{ij} - \mathcal{O}_{ji}}{2}.
    \end{align}
\end{subequations}
For a scalar field $\phi$ on the momentum-space manifold, the torsion field is defined as
\begin{equation}
    \mathcal{T}_{i j k} \partial^{i} \phi
    \equiv
    [\tilde{\nabla}_{j} , \tilde{\nabla}_{ k}] \phi
    =
    2 \Tilde{\Gamma}_{i[j k]}
    =
    2 C_{i[j k]}.
\end{equation}
Using these definitions, it is straightforward to verify that the connection difference tensor field $C$ naturally decomposes into the contorsion $K$ and disformation $L$ tensors as
\begin{equation}
C_{i j k}
=
K_{i j k} + L_{i j k},
\end{equation}
where
\begin{subequations}
\begin{align}
    K_{i j k}
    &= \frac{1}{2}
    (\mathcal{T}_{i j k}
    -
    \mathcal{T}_{j i k}
    -
    \mathcal{T}_{k i j}),
    \\
    L_{i j k}
    &= \frac{1}{2}
    (N_{i j k}
    +
    N_{k i j}
    -
    N_{j k i}).
\end{align}    
\end{subequations}
These inherit the symmetry properties of their constituents
\begin{subequations}
\begin{align}
    K_{i j k}
    &=
    K_{[i j] k},
    \\
    L_{i j k}
    &=
    L_{i(j k)}.
\end{align}    
\end{subequations}

Consider now a Hermitian connection component on the Bloch-state manifold
\begin{equation}
\label{SM_Eq_Hermitian_connection}
    \begin{split}
        \tilde{\Gamma}_{i j k}^{ba}
        &=
        A_{i}^{ab} (\mathcal{D}_{j} A_{k})^{ba}
        =
        A_{i}^{ab} \left(
        \partial_{j} A_{k}^{ba}
        -
        \text{i} [\mathcal{A}_{j} + A_{j}, A_{k}]^{ba}
        \right),
    \end{split}
\end{equation}
where $\mathcal{D}_j$ is the Berry covariant derivative~\cite{mehraeen2024quantum}, which takes into account the noncommutativity of the position operator and local momentum-space operators~\cite{blount1962formalisms}.
The first term on the right-hand side of Eq.~(\ref{SM_Eq_Hermitian_connection}) is reexpressed as
\begin{equation}
    \begin{split}
        A_{i}^{ab} \partial_{j} A_{k}^{ba}
        =&
        \frac{1}{2}
        \left[
        A_{i}^{ab} \partial_{j} A_{k}^{ba}
        +
        \partial_{j} ( A_{i}^{ab} A_{k}^{ba} )
        -
        \partial_{j} A_{i}^{ab} A_{k}^{ba} 
        \right]
        \\
        =&
        \frac{1}{2} \bigg\{
        \partial_{j} ( A_{i}^{ab} A_{k}^{ba} )
        +
        \partial_{k} ( A_{i}^{ab} A_{j}^{ba} )
        -
        \partial_{i} ( A_{j}^{ab} A_{k}^{ba} )
        +
        A_{j}^{ab} \partial_{i} A_{k}^{ba}
        -
        A_{j}^{ba} \partial_{i} A_{k}^{ab}
        \\
        &-
        A_{j}^{ba}
        \left[
        - \text{i} (\mathcal{A}_{i}^a - \mathcal{A}_{i}^b) A_{k}^{ab}
        + \text{i} (\mathcal{A}_{k}^a - \mathcal{A}_{k}^b) A_{i}^{ab}
        - \text{i} \sum_c (A_{i}^{ac} A_{k}^{cb}
        - A_{k}^{ac} A_{i}^{cb})
        \right]
        \\
        &+
        A_{i}^{ab}
        \left[
        - \text{i} (\mathcal{A}_{j}^a - \mathcal{A}_{j}^b) A_{k}^{ba}
        + \text{i} (\mathcal{A}_{k}^a - \mathcal{A}_{k}^b) A_{j}^{ba}
        + \text{i} \sum_c (A_{j}^{bc} A_{k}^{ca}
        - A_{k}^{bc} A_{j}^{ca})
        \right]
        \\
        &-
        A_{k}^{ba}
        \left[
        - \text{i} (\mathcal{A}_{i}^a - \mathcal{A}_{i}^b) A_{j}^{ab}
        + \text{i} (\mathcal{A}_{j}^a - \mathcal{A}_{j}^b) A_{i}^{ab}
        - \text{i} \sum_c (A_{i}^{ac} A_{j}^{cb}
        - A_{j}^{ac} A_{i}^{cb})
        \right]
        \Bigg\},
    \end{split}
\end{equation}
using the relation
\begin{equation}
    \partial_{i} A_{j}^{ab}
    -
    \partial_{j} A_{i}^{ab}
    =
    \text{i} [\mathcal{A}_{i} + A_{i} , \mathcal{A}_{j} + A_{j}]^{ab}.
\end{equation}
On reinserting into Eq.~(\ref{SM_Eq_Hermitian_connection}), and rearranging terms, we arrive at the identity
\begin{equation}
    C_{ijk}^{ba}
    =
    K_{ijk}^{ba}
    + \frac{\text{i}}{2} (\Gamma_{\Omega})_{ijk}^{ba}
    + \text{i} \Im (C_{ijk}^{ba}
    + \mathcal{T}_{kij}^{ba}),
\end{equation}
with $(\Gamma_{\Omega})_{ijk}^{ba}
=
\frac{1}{2} (
\partial_{k} \Omega_{ij}^{ba} 
+
\partial_{j} \Omega_{ik}^{ba}
-
\partial_{i} \Omega_{jk}^{ba})$.
From this identity, one can readily conclude that the Hilbert-space nonmetricity and disformation tensors are purely imaginary
\begin{subequations}
    \begin{align}
        C_{(ij)k}^{ba}
        &=
        \text{i} \Im \left[ C_{(ij)k}^{ba} \right],
        \\
        N_{ijk}^{ba}
        &=
        \text{i} \Im \left[ N_{ijk}^{ba} \right],
        \\
        L_{ijk}^{ba}
        &=
        \text{i} \Im \left[ L_{ijk}^{ba} \right],
    \end{align}
\end{subequations}
and moreover~\cite{Hsu2023, Mehraeen2025},
\begin{equation}
\Re \left( \tilde{\Gamma}_{ijk}^{ba} \right)
=
( \Gamma_{\text{g}} )_{i j k}^{ba}
+
\Re \left( K_{i j k}^{ba} \right).
\end{equation}

In the main text, we adapt the above decompositions to demonstrate that the imaginary Hilbert-space nonmetricity tensor~($N$), defined over the parameter space of deformation fields, dominates the odd nonlinear acoustoelastic responses realized in the considered axion insulator models. 

\section{Derivation of the nonlinear response functions}
\label{app:response}

\begin{figure*}
    \centering
\includegraphics[width = \linewidth]{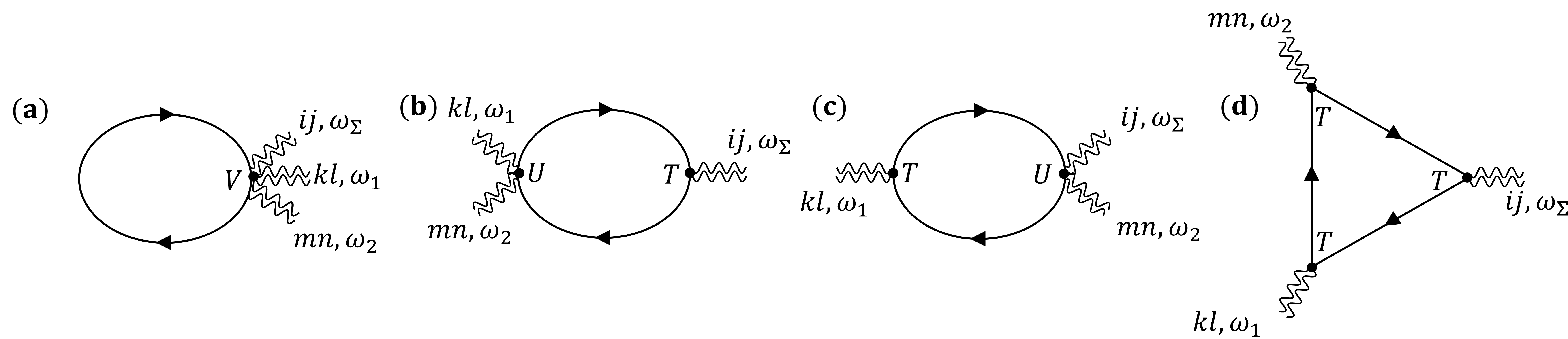}
    \caption{Diagrammatic structure of the quadratic acoustoelastic tensor $\eta_{ij;kl,mn}$, as a response to tensorial distortions $w_{kl}(\omega_1)$ and $w_{mn}(\omega_2)$. Diagrams {\bf(a)}-{\bf(c)} represent the bubble ($\medcirc$) contributions in Eq.~\eqref{eq:eta_circ}, and diagram {\bf(d)} is the triangle ($\triangle$) contribution in Eq.~\eqref{eq:eta_triangle}. Note that each diagram also implicitly denotes a conjugate diagram obtained by swapping $(kl,\omega_1)$ with $(mn, \omega_2)$, as required by intrinsic permutation symmetry of the second-order response tensor. Note further that {\bf(a)} is fully symmetric in the three index pairs.}
    \label{fig:diagrams}
\end{figure*}

The second-order response function $\eta_{B_0;B_1,B_2}$, for an observable $B_0$, upon perturbing fields $B_1$ and $B_2$, with frequencies $\omega_1$ and~$\omega_2$, respectively, can be expressed as the BZ integral of the $\vec{k}$-local response function. In this light, we work with $\vec{k}$-local quantities in this section. For a system with Hamiltonian $\mathcal{H}$, in~the presence of perturbing distortion fields $w_{ij} = \partial u_i/\partial x_j$, a comprehensive evaluation of the nonlinear elastic response tensor $\eta_{ij;kl,mn}$ requires considering variations of the Hamiltonian to third order in the distortion field
\begin{subequations}
    \begin{align}
        T_{ij}
        &=
        \frac{\delta \mathcal{H}}{\delta w_{ij}},
        \\
        U_{ij,kl}
        &=
        \frac{\delta^2 \mathcal{H}}
        {\delta w_{ij} \delta w_{kl}},
        \\
        V_{ij,kl,mn}
        &=
        \frac{\delta^3 \mathcal{H}}
        {\delta w_{ij} \delta w_{kl} \delta w_{mn}},
    \end{align}
\end{subequations}
where $T$ is the stress tensor operator and $U$ and $V$ are higher-order counterparts. Generalizing diagrammatic perturbation theory~\cite{Mahan2013, Parker2019, Rostami2021, Mehraeen2025} to higher-rank tensor distortion field perturbations (see Fig.~\ref{fig:diagrams}), the nonlinear elasticity tensor can be decomposed into bubble- and triangle-diagram contributions 
\begin{equation}
\label{eq:eta_decomp}
    \eta_{ij;kl,mn} = \eta^\medcirc_{ij;kl,mn}+ \eta^\triangle_{ij;kl,mn},
\end{equation}
which take the clean-limit forms~\cite{Bhalla2022,Bhalla2024}
\begin{align}
    \eta_{ij;kl,mn}^\medcirc(\omega_1^+, \omega_2^+) &= \lim_{\epsilon_1,\epsilon_2 \rightarrow 0}\frac{\text{i} (\omega_1^++\omega_2^+)}{(\text{i} \omega_1^+)(\text{i} \omega_2^+)}\sum_\mathcal{P} \sum_{a,b,c} \bigg( V^{aa}_{ij;kl,mn} f_a + f_{ab} \bigg[ \frac{T_{ij}^{ab} U_{kl,mn}^{ba}}{\omega_1^+ + \Delta^{ab}} + \frac{1}{2} \frac{T_{kl}^{ba} U_{ij,mn}^{ab} + T_{mn}^{ba} U_{ij,kl}^{ab}}{\omega_1^+ + \omega_2^+ + \Delta^{ab}}\bigg]\bigg), \label{eq:eta_circ}
    \\
    \eta_{ij;kl,mn}^\triangle(\omega_1^+, \omega_2^+)
    &= \lim_{\epsilon_1,\epsilon_2 \rightarrow 0}\frac{-\text{i} (\omega_1^++\omega_2^+)}{(\text{i} \omega_1^+)(\text{i} \omega_2^+)}\sum_{\mathcal{P}}\sum_{a,b,c}\frac{T_{ij}^{ab} T_{kl}^{bc} T_{mn}^{ca}}{\omega_1^+ + \omega_2^+ + \Delta^{ba}} \bigg ( \frac{f_{bc}}{\omega_1^+ + \Delta^{bc}}-\frac{f_{ca}}{\omega_2^+ + \Delta^{ca}}\bigg ) \label{eq:eta_triangle}.
\end{align}
%where $\Delta^{ab} = E^a - E^b$ represents band energy differences. 
Here, momentum labels are suppressed, $\sum_{\mathcal{P}}$ denotes internal permutation symmetrization between $(kl, \omega_1)$ and $(mn, \omega_2)$, band indices are denoted by $a, b, c$, coordinate indices by $i,j,k,l,m,n$, matrix elements $\mathcal{O}^{ab} = \bra{a}\mathcal{O}\ket{b}$ for any operator $\mathcal{O}$, $\Delta^{ab} = E^a - E^b$ represents band energy differences, and $f_{ab} = f_a - f_b$ with  $f_a$ denoting the Fermi filling factor for band $a$. Here and hereafter, we represent distortion variations $\delta / \delta w^{ij}$ as $\partial_{ij}$ for brevity. The $\medcirc$ term [Eq.~\eqref{eq:eta_circ}] is the contribution from the bubble diagrams [Fig.~\ref{fig:diagrams}(a)-(c)], while the $\triangle$ term [Eq.~\eqref{eq:eta_triangle}] accounts for the triangle diagram [Fig.~\ref{fig:diagrams}(d)]. In this work, we consider the odd (non-dissipative) part of the response tensor $\eta_{ij;kl,mn}$, which is given by~\cite{Jain2025NOVE}
\begin{equation}
\label{eq:odd_decomp}
    \eta^\mathrm{odd}_{ij;kl,mn} = \eta_{ij;kl,mn} - \eta_{(ij;kl,mn)}, 
\end{equation}
where, following Eq.~(\ref{SM_eq_symmetrization}), we note that (anti-)symmetrizations are henceforth understood to be over \textit{pairs} of coordinate indices, e.g., 
$\mathcal{O}_{(ij,kl)}
=
(\mathcal{O}_{ij,kl} + \mathcal{O}_{kl,ij})/2$.

We also note the following useful relations:
\begin{align}
    &T_{ij}^{ab} = \partial_{ij}E^b \delta^{ab} + \text{i} \Delta^{ab} A_{ij}^{ab}\,, \label{eq:T_ito_A}\\
    &U_{ij,kl}^{ab} = \partial_{ij}\partial_{kl}E^b\delta^{ab} + \text{i} \Delta^{ab}\partial_{ij} A_{kl}^{ab} +\text{i} (A_{ij}^{ab}\partial_{kl}\Delta^{ab} + A_{kl}^{ab}\partial_{ij}\Delta^{ab}) - \Delta^{ab} A_{kl}^{ab}(\mathcal{A}_{ij}^{b}-\mathcal{A}_{ij}^a)  -\sum_{c} (\Delta^{ac} A_{kl}^{ac}A_{ij}^{cb} + \Delta^{bc} A_{ij}^{ac}A_{ij}^{cb})\,,\label{eq:U_ito_A}\\
    &C_{ij;kl,mn}^{ab} =  A_{ij}^{ba}\partial_{kl}A_{mn}^{ab} + \text{i} Q_{ij,mn}^{ba}(\mathcal{A}_{kl}^{b}-\mathcal{A}_{kl}^a) + \mathcal{T}_{ij;kl,mn}^{ab}\,,\\
    &\mathcal{T}_{ij,kl,mn}^{ba} = -2\text{i} \sum_{c}Q_{ij,[kl,mn]}^{abc}\,, \label{eq:t_ito_Q}\\
    &N_{ij;kl,mn}^{ab} = 2C_{(ij;kl),mn}^{ab}.
\end{align}

We can now simplify each of the $\medcirc$ and $\triangle$ contributions to the odd part of the elastic response individually, beginning with the $\medcirc$ contribution. First, we note that the three-photon coupling [Fig.~\ref{fig:diagrams}(a), i.e., the first term on the right-hand side in Eq.~\eqref{eq:eta_circ}] is fully symmetric (dissipative) and does not contribute to the odd response. Next, we note that Eq.~\eqref{eq:eta_circ} is symmetric under $kl\leftrightarrow mn$, which allows us to expand the internal permutation symmetry explicitly as
\begin{equation}
    \eta_{ij;kl,mn}^\medcirc(\omega_1^+, \omega_2^+) = -\lim_{\epsilon_1,\epsilon_2 \rightarrow 0}\frac{\text{i}}{2} \frac{(\omega_1^+ + \omega_2^+)}{\omega_1^+ \omega_2^+}\sum_{ab}f_{ab} \bigg(T_{ij}^{ab} U_{kl,mn}^{ba} \bigg[ \frac{1}{\omega_1^+ + \Delta^{ab}} + \frac{1}{\omega_2^+ + \Delta^{ab}}\bigg] +\frac{T_{kl}^{ba} U_{ij,mn}^{ab} + T_{mn}^{ba} U_{ij,kl}^{ab}}{\omega_1^+ + \omega_2^+ + \Delta^{ab}}\bigg).
\end{equation}
Upon further index manipulations, we can write this as
\begin{equation}
    \eta_{ij;kl,mn}^\medcirc(\omega_1^+, \omega_2^+) = -\lim_{\epsilon_1,\epsilon_2 \rightarrow 0}\frac{\text{i}}{2} \frac{(\omega_1^+ + \omega_2^+)}{\omega_1^+ \omega_2^+}\sum_{ab}f_{ab}
    \left[T_{ij}^{ab} U_{kl,mn}^{ba} \beta_1^{ab} - (T_{kl}^{ab} U_{ij,mn}^{ba} + T_{mn}^{ab} U_{ij,kl}^{ba})\beta_2^{ab}
    \right],
\end{equation}
where $\beta_1^{ab}, \beta_2^{ab}$ are defined as,
\begin{align}
    \beta_1^{ab} &= \frac{1}{\omega_1^+ + \Delta^{ab}} + \frac{1}{\omega_2^+ + \Delta^{ab}}\,,\\
    \beta_2^{ab} &= \frac{1}{\omega_1^+ + \omega_2^+ + \Delta^{ab}}\,.\\
\end{align}
Next, we extract the odd part using Eq.~\eqref{eq:odd_decomp}, which yields
\begin{equation}
\label{eq:fgt}
    \eta_{ij;kl,mn}^{\medcirc\mathrm{,\,odd}}(\omega_1^+, \omega_2^+) = - \frac{\text{i}}{6}\sum_{ab}f_{ab}\underbrace{\lim_{\epsilon_1,\epsilon_2 \rightarrow 0}\bigg[\frac{(\omega_1^+ + \omega_2^+)(\beta_1^{ab}+\beta_2^{ab})}{\omega_1^+ \omega_2^+}\bigg]}_{\gamma^{ab}} \,\underbrace{\bigg[2T_{jk}^{ab}U_{kl,mn}^{ba}-T_{mn}^{ab}U_{ij,kl}^{ba}-T_{kl}^{ab}U_{ij,mn}^{ba}\bigg]}_{\theta_{ij;kl,mn}^{ab}}.
\end{equation}
We can now simplify the resonance structure factor $\gamma^{ab}$ and the matrix element contributions $\theta_{ij;kl,mn}^{ab}$ individually. Keeping in mind $\Delta^{ab} \neq 0$ (since we are working with gapped phases), and setting $\epsilon_1 = \epsilon_2$ in order to turn on the perturbing fields at the same rate~\cite{Rostami2021}, we can take the relevant limits to write
\begin{equation}
    \gamma^{ab} = \frac{2}{\Delta^{ab}}\begin{dcases} 
      \frac{1}{\text{i}\epsilon } - \frac{4}{\Delta^{ab}}\,, & \omega_i = 0 \,,\\
      \frac{1}{\omega} - \frac{4}{\Delta^{ab}} - \frac{2 \omega}{\Delta^{ab}} \,,& 0 < \frac{\omega_i}{\Delta^{ab}} \ll 1\,, \\
      0 \,,& \omega_1 + \omega_2 = 0.
   \end{dcases}
\end{equation}
Importantly, the dc limit must be taken by first setting $\omega_i=0$, followed by the small $\epsilon$ limit, while the non-zero frequency sub-gap acoustic SHG result is obtained by first taking the small $\epsilon$ limit, followed by an expansion in $\omega/\Delta^{ab}$. Note that the vanishing contribution to the rectification response is calculated at arbitrary frequency, and must be viewed in the sense of a~distributional identity, since the Sokhotsi-Plemelj theorem has been used. Notably, the absence of this contribution implies that the total rectification response must arise solely as a $\triangle$ contribution.

\begin{figure}[b!]
    \centering
\def\svgwidth{0.7\linewidth}
    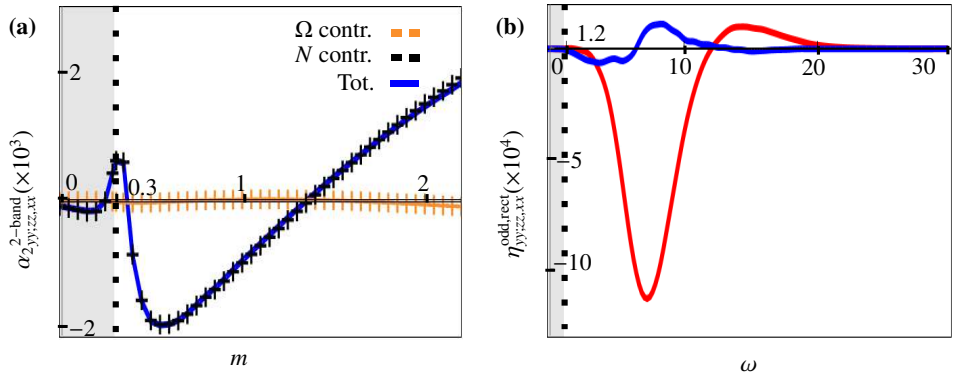
    \caption{Second independent component ($yy;zz,xx$) of the nonlinear odd acoustic responses of a model axion insulator~\cite{Wieder2018}. {\bf(a)} The $yy;zz,xx$ component of $\alpha_2^{\mathrm{2-band}}$ as a function of the spin-splitting mass parameter $m$. {\bf(b)} The $yy;zz,xx$ rectification response against frequency $\omega$.}
    \label{fig:SM2}
\end{figure}

On the other hand, using Eqs.~\eqref{eq:T_ito_A} and \eqref{eq:U_ito_A}, and enforcing symmetry under the index permutations, we find that $\theta_{ij;kl,mn}^{ab}$ simplifies to
\begin{align}
    \theta_{ij;kl,mn}^{ab}
    =& \Delta^{ab}
    \bigg\{
    2\bigg[2Q_{ij,(kl}^{ab}\partial_{mn)}\Delta^{ab}-Q_{kl,(ij}^{ab}\partial_{mn)}\Delta^{ab}-Q_{mn,(ij}^{ab}\partial_{kl)}\Delta^{ab}\bigg] + \Delta^{ab}\bigg[2C_{ij,(kl,mn)}^{ba}-C_{kl,(ij,mn)}^{ba}-C_{mn,(ij,kl)}^{ba}\bigg]\nonumber
    \\ 
    &+
    \text{i}\sum_{c}(\Delta^{ca}+ \Delta^{cb})
    \left[2Q_{ij,(kl,mn)}^{abc}-Q_{kl,(ij,mn)}^{abc}-Q_{mn,(ij,kl)}^{abc}
    \right]
    \bigg\}.
\end{align}
Due to the presence of $f_{ab}$ in Eq.~\eqref{eq:fgt}, the terms proportional to $\delta^{ab}$ in $\theta_{ij;kl,mn}^{ab}$ have been abandoned. Furthermore, only certain combinations of terms in the product $\gamma^{ab} \theta_{ij;kl,mn}^{ab}$ survive. In particular, we note that the $1/\text{i}\epsilon$, $1/\omega$ and $\omega$-linear terms in the expansion will combine non-trivially only with the $a \leftrightarrow b$ odd terms in $\theta$, while the $\omega$-independent terms $4/\Delta^{ab}$ are senstitive only to the even counterparts. This allows us to decompose the responses, as in the main text [Eqs.~(1) and (2)], as
\begin{align}
    \alpha_1^\medcirc
    =&
    \frac{2\text{i}}{3} \sum_{ab}f_{ba}\bigg(g_{(kl,mn)}^{ba}\partial_{ij}\Delta^{ba} - g_{ij,(kl}^{ba}\partial_{mn)}\Delta^{ba}\bigg)\,,
    \\
    \alpha_2^\medcirc
    =&
    4~\Bigg[ \sum_{ab}f_{ba} \bigg(\frac{\Omega_{ij,(kl}^{ba}\partial_{mn)}\Delta^{ba}}{\Delta^{ba}}+\frac{4\text{i}}{3}(N_{(kl,mn),ij}^{ba} - N_{ij;(kl,mn)}^{ba})\bigg) \nonumber 
    \\
    &+
    \frac{2}{3}\sum_{\substack{a,b,c
    \\
    \neq}}\frac{\Re Q_{ij,(kl,mn)}^{abc}}{\Delta^{ba}}\bigg( \frac{f_{bc}}{\Delta^{bc}}\bigg[2(\Delta^{bc})^2 - \Delta^{bc}\Delta^{ca} - (\Delta^{ca})^2\bigg] - \frac{f_{ca}}{\Delta^{ca}}\bigg[2(\Delta^{ca})^2 - \Delta^{ca}\Delta^{bc} - (\Delta^{bc})^2\bigg]\bigg)\Bigg]\,,
    \\
    \alpha_3^\medcirc
    =&
    \frac{4\text{i}}{3}~\Bigg[ \sum_{ab}\frac{f_{ba}}{\Delta^{ba}} \bigg(\frac{g_{(kl,mn)}^{ba}\partial_{ij}\Delta^{ba} - g_{ij,(kl}^{ba}\partial_{mn)}\Delta^{ba}}{\Delta^{ba}}\bigg) \nonumber
    \\
    &+\sum_{a,b,c}\frac{\Im Q_{ij,(kl,mn)}^{abc}}{(\Delta^{ba})^2}\bigg( \frac{f_{bc}\Delta^{ca}}{(\Delta^{bc})^2}\bigg[-2(\Delta^{bc})^2 + \Delta^{bc}\Delta^{ca} + (\Delta^{ca})^2\bigg] - \frac{f_{ca}\Delta^{bc}}{(\Delta^{ca})^2}\bigg[-2(\Delta^{ca})^2 + \Delta^{ca}\Delta^{bc} + (\Delta^{bc})^2\bigg]\bigg)\Bigg],
\end{align}
where we have split the above expressions into two- and three-band contributions. We have absorbed the torsion contributions into the three-band term using Eq.~\eqref{eq:t_ito_Q}, noting that the torsion and the three-state QGT are fundamentally three-band objects which vanish in two-band systems~\cite{Ahn2021, Mehraeen2025}.

We follow an analogous process for the $\triangle$ contributions. We take the appropriate sub-gap SHG limit and~extract the odd components to obtain
\begin{align}
    \alpha_1^\triangle
    =& 0,
    \\
    \alpha_2^\triangle
    =&
    3\sum_{ab}f_{ba}\frac{\Omega_{ij,(kl}^{ba}\partial_{mn)}\Delta^{ba}}{\Delta^{ba}} + 2 \sum_{\substack{a,b,c \\ \neq}} \frac{\Re Q_{ij,kl,mn}^{abc}}{\Delta^{ba}}\bigg( \frac{f_{bc}\Delta^{ca}}{\Delta^{bc}}[\Delta^{bc} + 3\Delta^{ca}] - \frac{f_{ca}\Delta^{bc}}{\Delta^{ca}}[\Delta^{ca} + 3\Delta^{bc}]\bigg),
    \\
    \alpha_3^\triangle
    =&
    2\text{i}~\bigg[ \sum_{ab}\frac{f_{ba}}{\Delta^{ba}} \bigg(\frac{g_{(kl,mn)}^{ba}\partial_{ij}\Delta^{ba} - g_{ij,(kl}^{ba}\partial_{mn)}\Delta^{ba}}{\Delta^{ba}}\bigg)
    \\ 
    &+
    \sum_{a,b,c}\frac{\Im Q_{ij,(kl,mn)}^{abc}}{(\Delta^{ba})^2}\bigg( \frac{f_{bc}\Delta^{ca}}{(\Delta^{bc})^2}[2(\Delta^{bc})^2 - \Delta^{bc}\Delta^{ca} -(\Delta^{ca})^2] - \frac{f_{ca}\Delta^{bc}}{(\Delta^{ca})^2}[2(\Delta^{ca})^2 - \Delta^{ca}\Delta^{bc} - (\Delta^{bc})^2]\bigg) \bigg].
\end{align}
It should be noted that the $\alpha_1$ and $\alpha_2$ components are consistent with the previous findings of Ref.~\cite{Jain2025NOVE}. On combining the bubble and triangle contributions, the total odd response is obtained as $\alpha_i = \alpha_i^\medcirc+\alpha_i^\triangle$. 

Finally, we obtain the rectification response by setting $\omega_1 = -\omega_2 = \omega$, followed by taking the limit $\epsilon \rightarrow 0$,
\begin{align}
    \eta_{ij;kl,mn}^{\mathrm{odd,rect}}(\omega,-\omega, \vec{k}) =& \pi \sum_{ab}f_{ab} \bigg[\frac{1}{3} \bigg(\delta(\Delta^{ba}-\omega) + \delta(\Delta^{ba}+\omega)\bigg)\bigg(g_{(kl,mn)}^{ba}\partial_{ij}\Delta^{ba} - g_{ij,(kl}^{ba}\partial_{mn)}\Delta^{ba}\bigg) \nonumber\\
    &+
    \frac{\text{i}}{4}\bigg(\delta(\Delta^{ba}-\omega) - \delta(\Delta^{ba}+\omega)\bigg) \bigg(\Omega_{kl,mn}^{ba}\partial_{ij}\Delta^{ba}\bigg) \bigg].
\end{align}
This completes the derivation of Eqs.~(1--6) central to the main text after integrating the $\vec{k}$-local quantities over the BZ.

\section{Nonlinear response coefficients}
In this Section, we present the response coefficients complementary to those highlighted in Fig.~2 of the main text. We note that the $yy;zz,xx$ component of $\alpha_2^{\mathrm{2-band}}$, plotted in Fig.~\ref{fig:SM2}{(a)}, shows again that the nonmetricity ($N$) contribution dominates over the Berry curvature ($\Omega$) contribution. We observe a sharp behaviour change at $m=M$, when the lower two bands become degenerate at $\Gamma$, similar to Fig.~2(a) in the main text. The behaviour of the rectification response [shown in Fig.~\ref{fig:SM2}(b)] is also qualitatively similar to the one demonstrated in the main text, Fig.~2(b). Furthermore, we note that the third coefficient ($zz;xx,yy$) can be inferred from the sum rule: $\eta_{xx;yy,zz}+\eta_{yy;zz,xx}+\eta_{zz;xx,yy}=0$~\cite{Jain2025NOVE}.

\begin{figure*}
    \centering
\def\svgwidth{\linewidth}
    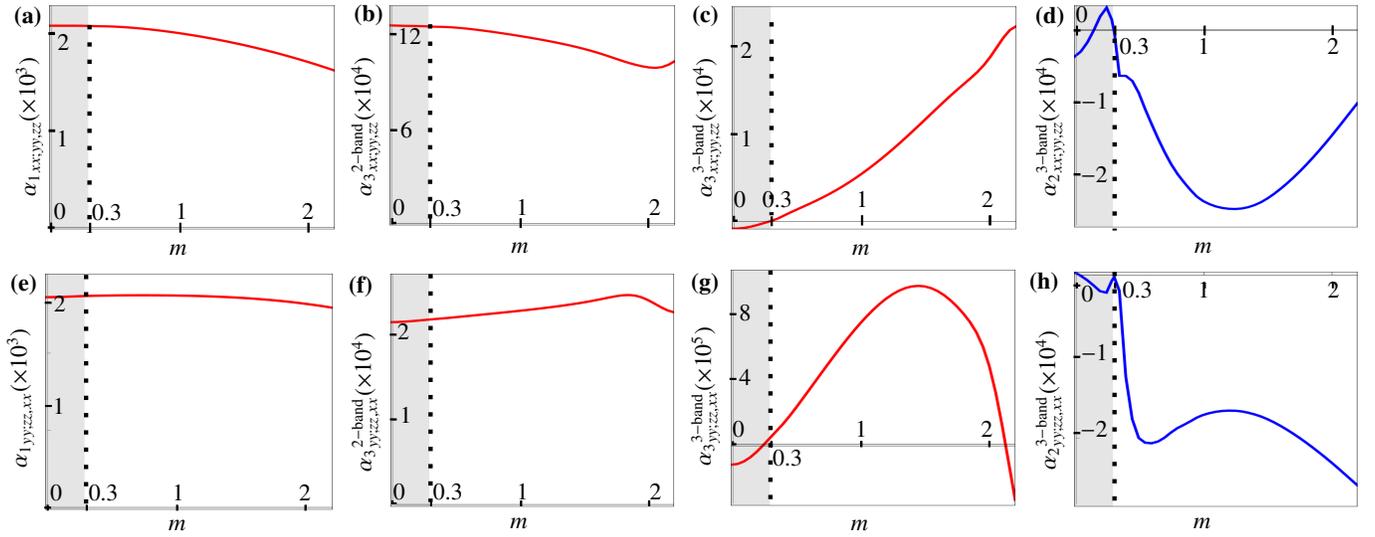
    \caption{Plots of $\alpha_1$, $\alpha_2^{\mathrm{3-band}}$ and $\alpha_3$ response terms (see Table~1 in the main text). We show both $xx;yy,zz$ and $yy;zz,xx$ response coefficients, complementary to the plots of $\alpha_2^{\mathrm{2-band}}$ presented in the main text (Fig.~2) and in Fig.~\ref{fig:SM2} above.}
    \label{fig:SM3}
\end{figure*}

In Fig.~\ref{fig:SM3}, we show the remaining SHG coefficients, as presented in Table~I of the main text, for both $xx;yy,zz$ and $yy;zz,xx$ responses. The real parts of the response [$\alpha_2^{\mathrm{3-band}}$, Figs.~\ref{fig:SM3}(d) and (h)] are again sensitive to the $m=M$ point, while the $\alpha_3^{\mathrm{3-band}}$ response [Figs.~\ref{fig:SM3}(c) and (g)] vanishes near $m=M$. 

\newpage
\bibliography{refs}

%% file: img/Fig1.pdf_tex
%% Creator: Inkscape 1.4.2 (f4327f4, 2025-05-13), www.inkscape.org
%% Accompanies image file 'Fig1.pdf' (pdf, eps, ps)
\begingroup%
  \makeatletter%
  \providecommand\color[2][]{%
    \errmessage{(Inkscape) Color is used for the text in Inkscape, but the package 'color.sty' is not loaded}%
    \renewcommand\color[2][]{}%
  }%
  \providecommand\transparent[1]{%
    \errmessage{(Inkscape) Transparency is used (non-zero) for the text in Inkscape, but the package 'transparent.sty' is not loaded}%
    \renewcommand\transparent[1]{}%
  }%
  \providecommand\rotatebox[2]{#2}%
  \newcommand*\fsize{\dimexpr\f@size pt\relax}%
  \newcommand*\lineheight[1]{\fontsize{\fsize}{#1\fsize}\selectfont}%
  \ifx\svgwidth\undefined%
    \setlength{\unitlength}{733.19181187bp}%
    \ifx\svgscale\undefined%
      \relax%
    \else%
      \setlength{\unitlength}{\unitlength * \real{\svgscale}}%
    \fi%
  \else%
    \setlength{\unitlength}{\svgwidth}%
  \fi%
  \global\let\svgwidth\undefined%
  \global\let\svgscale\undefined%
  \makeatother%
  \begin{picture}(1,0.38679409)%
    \lineheight{1}%
    \setlength\tabcolsep{0pt}%
    \put(0,0){\includegraphics[width=\unitlength,page=1]{Fig1.pdf}}%
    \put(0.88632359,0.14541274){\color[rgb]{0,0,0}\rotatebox{0.42357759}{\makebox(0,0)[lt]{\lineheight{1.25}\smash{\begin{tabular}[t]{l}$y$\end{tabular}}}}}%
    \put(0,0){\includegraphics[width=\unitlength,page=2]{Fig1.pdf}}%
    \put(0.88363224,0.00254345){\color[rgb]{0,0,0}\rotatebox{0.42357765}{\makebox(0,0)[lt]{\lineheight{1.25}\smash{\begin{tabular}[t]{l}$z$\end{tabular}}}}}%
    \put(0.97757723,0.03867098){\color[rgb]{0,0,0}\rotatebox{0.42357765}{\makebox(0,0)[lt]{\lineheight{1.25}\smash{\begin{tabular}[t]{l}$x$\end{tabular}}}}}%
    \put(0,0){\includegraphics[width=\unitlength,page=3]{Fig1.pdf}}%
    \put(0.14489344,0.06279739){\color[rgb]{1,0,0}\makebox(0,0)[lt]{\lineheight{1.25}\smash{\begin{tabular}[t]{l}$J_{zz}(2\omega)$\end{tabular}}}}%
    \put(-0.00054241,0.33491709){\color[rgb]{0,0,0}\makebox(0,0)[lt]{\lineheight{1.25}\smash{\begin{tabular}[t]{l}$\mathbf{(a)}$\end{tabular}}}}%
    \put(0,0){\includegraphics[width=\unitlength,page=4]{Fig1.pdf}}%
    \put(0.59047307,0.06163487){\color[rgb]{1,0,0}\makebox(0,0)[lt]{\lineheight{1.25}\smash{\begin{tabular}[t]{l}$J_{zz}^\mathrm{dc}$\end{tabular}}}}%
    \put(0.2366637,0.36539874){\color[rgb]{0,0.08235294,1}\makebox(0,0)[lt]{\lineheight{1.25}\smash{\begin{tabular}[t]{l}$w_{yy}(\omega)$\end{tabular}}}}%
    \put(0.68870249,0.36587992){\color[rgb]{0,0.08235294,1}\makebox(0,0)[lt]{\lineheight{1.25}\smash{\begin{tabular}[t]{l}$w_{yy}(-\omega)$\end{tabular}}}}%
    \put(0.36835696,0.20701087){\color[rgb]{0,0,0}\makebox(0,0)[lt]{\lineheight{1.25}\smash{\begin{tabular}[t]{l}$w_{xx}(\omega)$\end{tabular}}}}%
    \put(0.81862323,0.20766366){\color[rgb]{0,0,0}\makebox(0,0)[lt]{\lineheight{1.25}\smash{\begin{tabular}[t]{l}$w_{xx}(\omega)$\end{tabular}}}}%
    \put(0.44885366,0.3347087){\color[rgb]{0,0,0}\makebox(0,0)[lt]{\lineheight{1.25}\smash{\begin{tabular}[t]{l}$\mathbf{(b)}$\end{tabular}}}}%
    \put(0,0){\includegraphics[width=\unitlength,page=5]{Fig1.pdf}}%
    \put(0.1112923,0.18134701){\color[rgb]{0,0,0}\makebox(0,0)[lt]{\lineheight{1.25}\smash{\begin{tabular}[t]{l}$\theta = \pi$\end{tabular}}}}%
    \put(0.56275032,0.18223613){\color[rgb]{0,0,0}\makebox(0,0)[lt]{\lineheight{1.25}\smash{\begin{tabular}[t]{l}$\theta = \pi$\end{tabular}}}}%
    \put(0,0){\includegraphics[width=\unitlength,page=6]{Fig1.pdf}}%
  \end{picture}%
\endgroup%

%% file: img/Fig2_v5.pdf_tex
%% Creator: Inkscape 1.4.2 (f4327f4, 2025-05-13), www.inkscape.org
%% Accompanies image file 'Fig2_v5.pdf' (pdf, eps, ps)
\begingroup%
  \makeatletter%
  \providecommand\color[2][]{%
    \errmessage{(Inkscape) Color is used for the text in Inkscape, but the package 'color.sty' is not loaded}%
    \renewcommand\color[2][]{}%
  }%
  \providecommand\transparent[1]{%
    \errmessage{(Inkscape) Transparency is used (non-zero) for the text in Inkscape, but the package 'transparent.sty' is not loaded}%
    \renewcommand\transparent[1]{}%
  }%
  \providecommand\rotatebox[2]{#2}%
  \newcommand*\fsize{\dimexpr\f@size pt\relax}%
  \newcommand*\lineheight[1]{\fontsize{\fsize}{#1\fsize}\selectfont}%
  \ifx\svgwidth\undefined%
    \setlength{\unitlength}{366.57367376bp}%
    \ifx\svgscale\undefined%
      \relax%
    \else%
      \setlength{\unitlength}{\unitlength * \real{\svgscale}}%
    \fi%
  \else%
    \setlength{\unitlength}{\svgwidth}%
  \fi%
  \global\let\svgwidth\undefined%
  \global\let\svgscale\undefined%
  \makeatother%
  \begin{picture}(1,0.36320578)%
    \lineheight{1}%
    \setlength\tabcolsep{0pt}%
    \put(0.51167358,0.33170767){\color[rgb]{0,0,0}\makebox(0,0)[lt]{\lineheight{1.25}\smash{\begin{tabular}[t]{l}$\textbf{(b)}$\end{tabular}}}}%
    \put(0.55256562,0.05160251){\color[rgb]{0,0,0}\rotatebox{89.89116598}{\makebox(0,0)[lt]{\lineheight{1.25}\smash{\begin{tabular}[t]{l}$\eta^\mathrm{odd,rect}_{xx;yy,zz}(\times 10^4)$\end{tabular}}}}}%
    \put(0,0){\includegraphics[width=\unitlength,page=1]{Fig2_v5.pdf}}%
    \put(0.60313782,0.32021211){\color[rgb]{0,0,0}\makebox(0,0)[lt]{\lineheight{1.25}\smash{\begin{tabular}[t]{l}$1.2$\end{tabular}}}}%
    \put(0.58157809,0.26822643){\color[rgb]{0,0,0}\makebox(0,0)[lt]{\lineheight{1.25}\smash{\begin{tabular}[t]{l}$0$\end{tabular}}}}%
    \put(0.59200611,0.18404176){\color[rgb]{0,0,0}\makebox(0,0)[lt]{\lineheight{1.25}\smash{\begin{tabular}[t]{l}$-4$\end{tabular}}}}%
    \put(0.58975286,0.07019349){\color[rgb]{0,0,0}\makebox(0,0)[lt]{\lineheight{1.25}\smash{\begin{tabular}[t]{l}$-8$\end{tabular}}}}%
    \put(0,0){\includegraphics[width=\unitlength,page=2]{Fig2_v5.pdf}}%
    \put(0.70663167,0.31809108){\color[rgb]{0,0,0}\makebox(0,0)[lt]{\lineheight{1.25}\smash{\begin{tabular}[t]{l}$10$\end{tabular}}}}%
    \put(0.84532801,0.31749685){\color[rgb]{0,0,0}\makebox(0,0)[lt]{\lineheight{1.25}\smash{\begin{tabular}[t]{l}$20$\end{tabular}}}}%
    \put(0.9596754,0.31823204){\color[rgb]{0,0,0}\makebox(0,0)[lt]{\lineheight{1.25}\smash{\begin{tabular}[t]{l}$30$\end{tabular}}}}%
    \put(0.76492847,0.00535167){\color[rgb]{0,0,0}\makebox(0,0)[lt]{\lineheight{1.25}\smash{\begin{tabular}[t]{l}$\omega$\end{tabular}}}}%
    \put(0,0){\includegraphics[width=\unitlength,page=3]{Fig2_v5.pdf}}%
    \put(0.00015044,0.33550461){\color[rgb]{0,0,0}\makebox(0,0)[lt]{\lineheight{1.25}\smash{\begin{tabular}[t]{l}$\textbf{(a)}$\end{tabular}}}}%
    \put(0.03803278,0.04708955){\color[rgb]{0,0,0}\rotatebox{89.90598363}{\makebox(0,0)[lt]{\lineheight{1.25}\smash{\begin{tabular}[t]{l}${\alpha_{2}}_{xx;yy,zz}^\mathrm{2-band} (\times 10^3)$\end{tabular}}}}}%
    \put(0.35870252,0.16951482){\color[rgb]{0,0,0}\makebox(0,0)[lt]{\lineheight{1.25}\smash{\begin{tabular}[t]{l}Tot.\end{tabular}}}}%
    \put(0.28608665,0.22220268){\color[rgb]{0,0,0}\makebox(0,0)[lt]{\lineheight{1.25}\smash{\begin{tabular}[t]{l}$N$ contr.\end{tabular}}}}%
    \put(0.29475549,0.26955635){\color[rgb]{0,0,0}\makebox(0,0)[lt]{\lineheight{1.25}\smash{\begin{tabular}[t]{l}$\Omega$ contr.\end{tabular}}}}%
    \put(0,0){\includegraphics[width=\unitlength,page=4]{Fig2_v5.pdf}}%
    \put(0.07558459,0.27998131){\color[rgb]{0,0,0}\makebox(0,0)[lt]{\lineheight{1.25}\smash{\begin{tabular}[t]{l}$0$\end{tabular}}}}%
    \put(0.13100879,0.33013403){\color[rgb]{0,0,0}\makebox(0,0)[lt]{\lineheight{1.25}\smash{\begin{tabular}[t]{l}$0.3$\end{tabular}}}}%
    \put(0.24725395,0.33069172){\color[rgb]{0,0,0}\makebox(0,0)[lt]{\lineheight{1.25}\smash{\begin{tabular}[t]{l}$1$\end{tabular}}}}%
    \put(0.43621416,0.33069172){\color[rgb]{0,0,0}\makebox(0,0)[lt]{\lineheight{1.25}\smash{\begin{tabular}[t]{l}$2$\end{tabular}}}}%
    \put(0.07130415,0.17725659){\color[rgb]{0,0,0}\makebox(0,0)[lt]{\lineheight{1.25}\smash{\begin{tabular}[t]{l}$-4$\end{tabular}}}}%
    \put(0.07058046,0.05083759){\color[rgb]{0,0,0}\makebox(0,0)[lt]{\lineheight{1.25}\smash{\begin{tabular}[t]{l}$-8$\end{tabular}}}}%
    \put(0.24281254,0.00977398){\color[rgb]{0,0,0}\makebox(0,0)[lt]{\lineheight{1.25}\smash{\begin{tabular}[t]{l}$m$\end{tabular}}}}%
    \put(0,0){\includegraphics[width=\unitlength,page=5]{Fig2_v5.pdf}}%
  \end{picture}%
\endgroup%

%% file: img/FinalSM2_v2.pdf_tex
%% Creator: Inkscape 1.4.2 (f4327f4, 2025-05-13), www.inkscape.org
%% Accompanies image file 'FinalSM2_v2.pdf' (pdf, eps, ps)
\begingroup%
  \makeatletter%
  \providecommand\color[2][]{%
    \errmessage{(Inkscape) Color is used for the text in Inkscape, but the package 'color.sty' is not loaded}%
    \renewcommand\color[2][]{}%
  }%
  \providecommand\transparent[1]{%
    \errmessage{(Inkscape) Transparency is used (non-zero) for the text in Inkscape, but the package 'transparent.sty' is not loaded}%
    \renewcommand\transparent[1]{}%
  }%
  \providecommand\rotatebox[2]{#2}%
  \newcommand*\fsize{\dimexpr\f@size pt\relax}%
  \newcommand*\lineheight[1]{\fontsize{\fsize}{#1\fsize}\selectfont}%
  \ifx\svgwidth\undefined%
    \setlength{\unitlength}{348.36764118bp}%
    \ifx\svgscale\undefined%
      \relax%
    \else%
      \setlength{\unitlength}{\unitlength * \real{\svgscale}}%
    \fi%
  \else%
    \setlength{\unitlength}{\svgwidth}%
  \fi%
  \global\let\svgwidth\undefined%
  \global\let\svgscale\undefined%
  \makeatother%
  \begin{picture}(1,0.47152181)%
    \lineheight{1}%
    \setlength\tabcolsep{0pt}%
    \put(0.54121816,0.12299176){\color[rgb]{0,0,0}\rotatebox{89.87830578}{\makebox(0,0)[lt]{\lineheight{1.25}\smash{\begin{tabular}[t]{l}$\eta^\mathrm{odd,rect}_{yy;zz,xx}(\times 10^4)$\end{tabular}}}}}%
    \put(0,0){\includegraphics[width=\unitlength,page=1]{FinalSM2_v2.pdf}}%
    \put(0.02181185,0.11120685){\color[rgb]{0,0,0}\rotatebox{89.92932384}{\makebox(0,0)[lt]{\lineheight{1.25}\smash{\begin{tabular}[t]{l}${\alpha_{2}}_{yy;zz,xx}^\mathrm{2-band} (\times 10^3)$\end{tabular}}}}}%
    \put(0.51615907,0.36110612){\color[rgb]{0,0,0}\makebox(0,0)[lt]{\lineheight{1.25}\smash{\begin{tabular}[t]{l}$\textbf{(b)}$\end{tabular}}}}%
    \put(0.60158721,0.35375326){\color[rgb]{0,0,0}\makebox(0,0)[lt]{\lineheight{1.25}\smash{\begin{tabular}[t]{l}$1.2$\end{tabular}}}}%
    \put(0.57528348,0.31856427){\color[rgb]{0,0,0}\makebox(0,0)[lt]{\lineheight{1.25}\smash{\begin{tabular}[t]{l}$0$\end{tabular}}}}%
    \put(0.57983477,0.21826946){\color[rgb]{0,0,0}\makebox(0,0)[lt]{\lineheight{1.25}\smash{\begin{tabular}[t]{l}$-5$\end{tabular}}}}%
    \put(0.57672824,0.11003111){\color[rgb]{0,0,0}\makebox(0,0)[lt]{\lineheight{1.25}\smash{\begin{tabular}[t]{l}$-10$\end{tabular}}}}%
    \put(0,0){\includegraphics[width=\unitlength,page=2]{FinalSM2_v2.pdf}}%
    \put(0.69864887,0.31804164){\color[rgb]{0,0,0}\makebox(0,0)[lt]{\lineheight{1.25}\smash{\begin{tabular}[t]{l}$10$\end{tabular}}}}%
    \put(0.84037452,0.31771301){\color[rgb]{0,0,0}\makebox(0,0)[lt]{\lineheight{1.25}\smash{\begin{tabular}[t]{l}$20$\end{tabular}}}}%
    \put(0.95721958,0.31811949){\color[rgb]{0,0,0}\makebox(0,0)[lt]{\lineheight{1.25}\smash{\begin{tabular}[t]{l}$30$\end{tabular}}}}%
    \put(0.77548995,0.00353831){\color[rgb]{0,0,0}\makebox(0,0)[lt]{\lineheight{1.25}\smash{\begin{tabular}[t]{l}$\omega$\end{tabular}}}}%
    \put(0,0){\includegraphics[width=\unitlength,page=3]{FinalSM2_v2.pdf}}%
    \put(0.000189,0.36031243){\color[rgb]{0,0,0}\makebox(0,0)[lt]{\lineheight{1.25}\smash{\begin{tabular}[t]{l}$\textbf{(a)}$\end{tabular}}}}%
    \put(0.34862787,0.30179428){\color[rgb]{0,0,0}\makebox(0,0)[lt]{\lineheight{1.25}\smash{\begin{tabular}[t]{l}Tot.\end{tabular}}}}%
    \put(0.30346198,0.33092768){\color[rgb]{0,0,0}\makebox(0,0)[lt]{\lineheight{1.25}\smash{\begin{tabular}[t]{l}$N$ contr.\end{tabular}}}}%
    \put(0.30369939,0.35711165){\color[rgb]{0,0,0}\makebox(0,0)[lt]{\lineheight{1.25}\smash{\begin{tabular}[t]{l}$\Omega$ contr.\end{tabular}}}}%
    \put(0,0){\includegraphics[width=\unitlength,page=4]{FinalSM2_v2.pdf}}%
    \put(0.06054547,0.19507526){\color[rgb]{0,0,0}\makebox(0,0)[lt]{\lineheight{1.25}\smash{\begin{tabular}[t]{l}$0$\end{tabular}}}}%
    \put(0.1266634,0.19026953){\color[rgb]{0,0,0}\makebox(0,0)[lt]{\lineheight{1.25}\smash{\begin{tabular}[t]{l}$0.3$\end{tabular}}}}%
    \put(0.24014605,0.19466833){\color[rgb]{0,0,0}\makebox(0,0)[lt]{\lineheight{1.25}\smash{\begin{tabular}[t]{l}$1$\end{tabular}}}}%
    \put(0.43323341,0.19466833){\color[rgb]{0,0,0}\makebox(0,0)[lt]{\lineheight{1.25}\smash{\begin{tabular}[t]{l}$2$\end{tabular}}}}%
    \put(0.06453416,0.31142805){\color[rgb]{0,0,0}\makebox(0,0)[lt]{\lineheight{1.25}\smash{\begin{tabular}[t]{l}$2$\end{tabular}}}}%
    \put(0.06379478,0.04031704){\color[rgb]{0,0,0}\makebox(0,0)[lt]{\lineheight{1.25}\smash{\begin{tabular}[t]{l}$-2$\end{tabular}}}}%
    \put(0.235302,0.00926493){\color[rgb]{0,0,0}\makebox(0,0)[lt]{\lineheight{1.25}\smash{\begin{tabular}[t]{l}$m$\end{tabular}}}}%
    \put(0,0){\includegraphics[width=\unitlength,page=5]{FinalSM2_v2.pdf}}%
  \end{picture}%
\endgroup%

%% file: img/FinalSM.pdf_tex
%% Creator: Inkscape 1.4.2 (f4327f4, 2025-05-13), www.inkscape.org
%% Accompanies image file 'FinalSM.pdf' (pdf, eps, ps)
\begingroup%
  \makeatletter%
  \providecommand\color[2][]{%
    \errmessage{(Inkscape) Color is used for the text in Inkscape, but the package 'color.sty' is not loaded}%
    \renewcommand\color[2][]{}%
  }%
  \providecommand\transparent[1]{%
    \errmessage{(Inkscape) Transparency is used (non-zero) for the text in Inkscape, but the package 'transparent.sty' is not loaded}%
    \renewcommand\transparent[1]{}%
  }%
  \providecommand\rotatebox[2]{#2}%
  \newcommand*\fsize{\dimexpr\f@size pt\relax}%
  \newcommand*\lineheight[1]{\fontsize{\fsize}{#1\fsize}\selectfont}%
  \ifx\svgwidth\undefined%
    \setlength{\unitlength}{708.36526225bp}%
    \ifx\svgscale\undefined%
      \relax%
    \else%
      \setlength{\unitlength}{\unitlength * \real{\svgscale}}%
    \fi%
  \else%
    \setlength{\unitlength}{\svgwidth}%
  \fi%
  \global\let\svgwidth\undefined%
  \global\let\svgscale\undefined%
  \makeatother%
  \begin{picture}(1,0.39175335)%
    \lineheight{1}%
    \setlength\tabcolsep{0pt}%
    \put(0,0){\includegraphics[width=\unitlength,page=1]{FinalSM.pdf}}%
    \put(0.76033503,0.37944744){\color[rgb]{0,0,0}\makebox(0,0)[lt]{\lineheight{1.25}\smash{\begin{tabular}[t]{l}$\textbf{(d)}$\end{tabular}}}}%
    \put(0.77539163,0.25052429){\color[rgb]{0,0,0}\rotatebox{90.26428109}{\makebox(0,0)[lt]{\lineheight{1.25}\smash{\begin{tabular}[t]{l}${\alpha_{2}}_{xx;yy,zz}^\mathrm{3-band} (\times 10^4)$\end{tabular}}}}}%
    \put(0.79043465,0.37958388){\color[rgb]{0,0,0}\makebox(0,0)[lt]{\lineheight{1.25}\smash{\begin{tabular}[t]{l}$0$\end{tabular}}}}%
    \put(0.82281643,0.35615976){\color[rgb]{0,0,0}\makebox(0,0)[lt]{\lineheight{1.25}\smash{\begin{tabular}[t]{l}$0.3$\end{tabular}}}}%
    \put(0.88080512,0.35644836){\color[rgb]{0,0,0}\makebox(0,0)[lt]{\lineheight{1.25}\smash{\begin{tabular}[t]{l}$1$\end{tabular}}}}%
    \put(0.97667986,0.35644836){\color[rgb]{0,0,0}\makebox(0,0)[lt]{\lineheight{1.25}\smash{\begin{tabular}[t]{l}$2$\end{tabular}}}}%
    \put(0,0){\includegraphics[width=\unitlength,page=2]{FinalSM.pdf}}%
    \put(0.79436234,0.31577543){\color[rgb]{0,0,0}\makebox(0,0)[lt]{\lineheight{1.25}\smash{\begin{tabular}[t]{l}$-1$\end{tabular}}}}%
    \put(0.79395243,0.26084162){\color[rgb]{0,0,0}\makebox(0,0)[lt]{\lineheight{1.25}\smash{\begin{tabular}[t]{l}$-2$\end{tabular}}}}%
    \put(0.87836952,0.20838485){\color[rgb]{0,0,0}\makebox(0,0)[lt]{\lineheight{1.25}\smash{\begin{tabular}[t]{l}$m$\end{tabular}}}}%
    \put(0,0){\includegraphics[width=\unitlength,page=3]{FinalSM.pdf}}%
    \put(0.50583501,0.37987753){\color[rgb]{0,0,0}\makebox(0,0)[lt]{\lineheight{1.25}\smash{\begin{tabular}[t]{l}$\textbf{(c)}$\end{tabular}}}}%
    \put(0.52031665,0.25007647){\color[rgb]{0,0,0}\rotatebox{89.74262704}{\makebox(0,0)[lt]{\lineheight{1.25}\smash{\begin{tabular}[t]{l}${\alpha_{3}}_{xx;yy,zz}^\mathrm{3-band} (\times 10^4)$\end{tabular}}}}}%
    \put(0.7204292,0.24274436){\color[rgb]{0,0,0}\makebox(0,0)[lt]{\lineheight{1.25}\smash{\begin{tabular}[t]{l}$2$\end{tabular}}}}%
    \put(0.62734601,0.24274436){\color[rgb]{0,0,0}\makebox(0,0)[lt]{\lineheight{1.25}\smash{\begin{tabular}[t]{l}$1$\end{tabular}}}}%
    \put(0.53660857,0.24287474){\color[rgb]{0,0,0}\makebox(0,0)[lt]{\lineheight{1.25}\smash{\begin{tabular}[t]{l}$0$\end{tabular}}}}%
    \put(0.54150802,0.28738332){\color[rgb]{0,0,0}\makebox(0,0)[lt]{\lineheight{1.25}\smash{\begin{tabular}[t]{l}$1$\end{tabular}}}}%
    \put(0.54150802,0.35302462){\color[rgb]{0,0,0}\makebox(0,0)[lt]{\lineheight{1.25}\smash{\begin{tabular}[t]{l}$2$\end{tabular}}}}%
    \put(0.55784139,0.24220579){\color[rgb]{0,0,0}\makebox(0,0)[lt]{\lineheight{1.25}\smash{\begin{tabular}[t]{l}$0.3$\end{tabular}}}}%
    \put(0.62442459,0.20719203){\color[rgb]{0,0,0}\makebox(0,0)[lt]{\lineheight{1.25}\smash{\begin{tabular}[t]{l}$m$\end{tabular}}}}%
    \put(0,0){\includegraphics[width=\unitlength,page=4]{FinalSM.pdf}}%
    \put(0.2551015,0.38047007){\color[rgb]{0,0,0}\makebox(0,0)[lt]{\lineheight{1.25}\smash{\begin{tabular}[t]{l}$\textbf{(b)}$\end{tabular}}}}%
    \put(0.2676803,0.25092962){\color[rgb]{0,0,0}\rotatebox{90.27525034}{\makebox(0,0)[lt]{\lineheight{1.25}\smash{\begin{tabular}[t]{l}${\alpha_{3}}_{xx;yy,zz}^\mathrm{2-band} (\times 10^4)$\end{tabular}}}}}%
    \put(0.28373052,0.23674736){\color[rgb]{0,0,0}\makebox(0,0)[lt]{\lineheight{1.25}\smash{\begin{tabular}[t]{l}$0$\end{tabular}}}}%
    \put(0.3132099,0.23607829){\color[rgb]{0,0,0}\makebox(0,0)[lt]{\lineheight{1.25}\smash{\begin{tabular}[t]{l}$0.3$\end{tabular}}}}%
    \put(0.37521375,0.23636698){\color[rgb]{0,0,0}\makebox(0,0)[lt]{\lineheight{1.25}\smash{\begin{tabular}[t]{l}$1$\end{tabular}}}}%
    \put(0.47360147,0.23679518){\color[rgb]{0,0,0}\makebox(0,0)[lt]{\lineheight{1.25}\smash{\begin{tabular}[t]{l}$2$\end{tabular}}}}%
    \put(0.28932953,0.29420382){\color[rgb]{0,0,0}\makebox(0,0)[lt]{\lineheight{1.25}\smash{\begin{tabular}[t]{l}$6$\end{tabular}}}}%
    \put(0.28725232,0.36386639){\color[rgb]{0,0,0}\makebox(0,0)[lt]{\lineheight{1.25}\smash{\begin{tabular}[t]{l}$12$\end{tabular}}}}%
    \put(0.37159766,0.20878521){\color[rgb]{0,0,0}\makebox(0,0)[lt]{\lineheight{1.25}\smash{\begin{tabular}[t]{l}$m$\end{tabular}}}}%
    \put(0,0){\includegraphics[width=\unitlength,page=5]{FinalSM.pdf}}%
    \put(0.00274346,0.37917975){\color[rgb]{0,0,0}\makebox(0,0)[lt]{\lineheight{1.25}\smash{\begin{tabular}[t]{l}$\textbf{(a)}$\end{tabular}}}}%
    \put(0.01879603,0.2505461){\color[rgb]{0,0,0}\rotatebox{90.11397247}{\makebox(0,0)[lt]{\lineheight{1.25}\smash{\begin{tabular}[t]{l}${\alpha_{1}}_{xx;yy,zz} (\times 10^3)$\end{tabular}}}}}%
    \put(0.0322672,0.23396716){\color[rgb]{0,0,0}\makebox(0,0)[lt]{\lineheight{1.25}\smash{\begin{tabular}[t]{l}$0$\end{tabular}}}}%
    \put(0.06063165,0.23329825){\color[rgb]{0,0,0}\makebox(0,0)[lt]{\lineheight{1.25}\smash{\begin{tabular}[t]{l}$0.3$\end{tabular}}}}%
    \put(0.12263557,0.23358686){\color[rgb]{0,0,0}\makebox(0,0)[lt]{\lineheight{1.25}\smash{\begin{tabular}[t]{l}$1$\end{tabular}}}}%
    \put(0.21627835,0.23358659){\color[rgb]{0,0,0}\makebox(0,0)[lt]{\lineheight{1.25}\smash{\begin{tabular}[t]{l}$2$\end{tabular}}}}%
    \put(0.11830976,0.20685487){\color[rgb]{0,0,0}\makebox(0,0)[lt]{\lineheight{1.25}\smash{\begin{tabular}[t]{l}$m$\end{tabular}}}}%
    \put(0.87758469,0.00398152){\color[rgb]{0,0,0}\makebox(0,0)[lt]{\lineheight{1.25}\smash{\begin{tabular}[t]{l}$m$\end{tabular}}}}%
    \put(0.62363995,0.00278873){\color[rgb]{0,0,0}\makebox(0,0)[lt]{\lineheight{1.25}\smash{\begin{tabular}[t]{l}$m$\end{tabular}}}}%
    \put(0.37081297,0.00438207){\color[rgb]{0,0,0}\makebox(0,0)[lt]{\lineheight{1.25}\smash{\begin{tabular}[t]{l}$m$\end{tabular}}}}%
    \put(0.11752506,0.0024515){\color[rgb]{0,0,0}\makebox(0,0)[lt]{\lineheight{1.25}\smash{\begin{tabular}[t]{l}$m$\end{tabular}}}}%
    \put(0,0){\includegraphics[width=\unitlength,page=6]{FinalSM.pdf}}%
    \put(0.03424001,0.28732127){\color[rgb]{0,0,0}\makebox(0,0)[lt]{\lineheight{1.25}\smash{\begin{tabular}[t]{l}$1$\end{tabular}}}}%
    \put(0,0){\includegraphics[width=\unitlength,page=7]{FinalSM.pdf}}%
    \put(0.03495827,0.3592587){\color[rgb]{0,0,0}\makebox(0,0)[lt]{\lineheight{1.25}\smash{\begin{tabular}[t]{l}$2$\end{tabular}}}}%
    \put(0,0){\includegraphics[width=\unitlength,page=8]{FinalSM.pdf}}%
    \put(0.75590853,0.18184315){\color[rgb]{0,0,0}\makebox(0,0)[lt]{\lineheight{1.25}\smash{\begin{tabular}[t]{l}$\textbf{(h)}$\end{tabular}}}}%
    \put(0.77523068,0.04770748){\color[rgb]{0,0,0}\rotatebox{90.19580988}{\makebox(0,0)[lt]{\lineheight{1.25}\smash{\begin{tabular}[t]{l}${\alpha_{2}}_{yy;zz,xx}^\mathrm{3-band} (\times 10^4)$\end{tabular}}}}}%
    \put(0.79454194,0.17282483){\color[rgb]{0,0,0}\makebox(0,0)[lt]{\lineheight{1.25}\smash{\begin{tabular}[t]{l}$0$\end{tabular}}}}%
    \put(0.82499111,0.17523263){\color[rgb]{0,0,0}\makebox(0,0)[lt]{\lineheight{1.25}\smash{\begin{tabular}[t]{l}$0.3$\end{tabular}}}}%
    \put(0.88086727,0.17538432){\color[rgb]{0,0,0}\makebox(0,0)[lt]{\lineheight{1.25}\smash{\begin{tabular}[t]{l}$1$\end{tabular}}}}%
    \put(0.97694337,0.17538432){\color[rgb]{0,0,0}\makebox(0,0)[lt]{\lineheight{1.25}\smash{\begin{tabular}[t]{l}$2$\end{tabular}}}}%
    \put(0,0){\includegraphics[width=\unitlength,page=9]{FinalSM.pdf}}%
    \put(0.79424282,0.12909176){\color[rgb]{0,0,0}\makebox(0,0)[lt]{\lineheight{1.25}\smash{\begin{tabular}[t]{l}$-1$\end{tabular}}}}%
    \put(0.79383206,0.07171188){\color[rgb]{0,0,0}\makebox(0,0)[lt]{\lineheight{1.25}\smash{\begin{tabular}[t]{l}$-2$\end{tabular}}}}%
    \put(0,0){\includegraphics[width=\unitlength,page=10]{FinalSM.pdf}}%
    \put(0.50494978,0.18137304){\color[rgb]{0,0,0}\makebox(0,0)[lt]{\lineheight{1.25}\smash{\begin{tabular}[t]{l}$\textbf{(g)}$\end{tabular}}}}%
    \put(0.51946313,0.05047142){\color[rgb]{0,0,0}\rotatebox{89.80930985}{\makebox(0,0)[lt]{\lineheight{1.25}\smash{\begin{tabular}[t]{l}${\alpha_{3}}_{yy;zz,xx}^\mathrm{3-band} (\times 10^5)$\end{tabular}}}}}%
    \put(0.72000192,0.0720098){\color[rgb]{0,0,0}\makebox(0,0)[lt]{\lineheight{1.25}\smash{\begin{tabular}[t]{l}$2$\end{tabular}}}}%
    \put(0.62672066,0.0720098){\color[rgb]{0,0,0}\makebox(0,0)[lt]{\lineheight{1.25}\smash{\begin{tabular}[t]{l}$1$\end{tabular}}}}%
    \put(0.53578915,0.07207828){\color[rgb]{0,0,0}\makebox(0,0)[lt]{\lineheight{1.25}\smash{\begin{tabular}[t]{l}$0$\end{tabular}}}}%
    \put(0.54069909,0.11018304){\color[rgb]{0,0,0}\makebox(0,0)[lt]{\lineheight{1.25}\smash{\begin{tabular}[t]{l}$4$\end{tabular}}}}%
    \put(0.54069909,0.15847271){\color[rgb]{0,0,0}\makebox(0,0)[lt]{\lineheight{1.25}\smash{\begin{tabular}[t]{l}$8$\end{tabular}}}}%
    \put(0.56542078,0.0505512){\color[rgb]{0,0,0}\makebox(0,0)[lt]{\lineheight{1.25}\smash{\begin{tabular}[t]{l}$0.3$\end{tabular}}}}%
    \put(0,0){\includegraphics[width=\unitlength,page=11]{FinalSM.pdf}}%
    \put(0.25104105,0.17943073){\color[rgb]{0,0,0}\makebox(0,0)[lt]{\lineheight{1.25}\smash{\begin{tabular}[t]{l}$\textbf{(f)}$\end{tabular}}}}%
    \put(0.26788105,0.04637288){\color[rgb]{0,0,0}\rotatebox{90.20393404}{\makebox(0,0)[lt]{\lineheight{1.25}\smash{\begin{tabular}[t]{l}${\alpha_{3}}_{yy;zz,xx}^\mathrm{2-band} (\times 10^4)$\end{tabular}}}}}%
    \put(0.28396612,0.0258942){\color[rgb]{0,0,0}\makebox(0,0)[lt]{\lineheight{1.25}\smash{\begin{tabular}[t]{l}$0$\end{tabular}}}}%
    \put(0.31348105,0.02554259){\color[rgb]{0,0,0}\makebox(0,0)[lt]{\lineheight{1.25}\smash{\begin{tabular}[t]{l}$0.3$\end{tabular}}}}%
    \put(0.37564351,0.02569424){\color[rgb]{0,0,0}\makebox(0,0)[lt]{\lineheight{1.25}\smash{\begin{tabular}[t]{l}$1$\end{tabular}}}}%
    \put(0.47424022,0.02591929){\color[rgb]{0,0,0}\makebox(0,0)[lt]{\lineheight{1.25}\smash{\begin{tabular}[t]{l}$2$\end{tabular}}}}%
    \put(0.2890243,0.08217766){\color[rgb]{0,0,0}\makebox(0,0)[lt]{\lineheight{1.25}\smash{\begin{tabular}[t]{l}$1$\end{tabular}}}}%
    \put(0.28731108,0.14415439){\color[rgb]{0,0,0}\makebox(0,0)[lt]{\lineheight{1.25}\smash{\begin{tabular}[t]{l}$2$\end{tabular}}}}%
    \put(0,0){\includegraphics[width=\unitlength,page=12]{FinalSM.pdf}}%
    \put(0.00009295,0.18062983){\color[rgb]{0,0,0}\makebox(0,0)[lt]{\lineheight{1.25}\smash{\begin{tabular}[t]{l}$\textbf{(e)}$\end{tabular}}}}%
    \put(0.01219699,0.04920338){\color[rgb]{0,0,0}\rotatebox{90.08444289}{\makebox(0,0)[lt]{\lineheight{1.25}\smash{\begin{tabular}[t]{l}${\alpha_{1}}_{yy;zz,xx} (\times 10^3)$\end{tabular}}}}}%
    \put(0.02967939,0.02519764){\color[rgb]{0,0,0}\makebox(0,0)[lt]{\lineheight{1.25}\smash{\begin{tabular}[t]{l}$0$\end{tabular}}}}%
    \put(0.05807716,0.0248461){\color[rgb]{0,0,0}\makebox(0,0)[lt]{\lineheight{1.25}\smash{\begin{tabular}[t]{l}$0.3$\end{tabular}}}}%
    \put(0.12023967,0.02499769){\color[rgb]{0,0,0}\makebox(0,0)[lt]{\lineheight{1.25}\smash{\begin{tabular}[t]{l}$1$\end{tabular}}}}%
    \put(0.2140814,0.02499769){\color[rgb]{0,0,0}\makebox(0,0)[lt]{\lineheight{1.25}\smash{\begin{tabular}[t]{l}$2$\end{tabular}}}}%
    \put(0.03165643,0.08872054){\color[rgb]{0,0,0}\makebox(0,0)[lt]{\lineheight{1.25}\smash{\begin{tabular}[t]{l}$1$\end{tabular}}}}%
    \put(0.03237622,0.16515396){\color[rgb]{0,0,0}\makebox(0,0)[lt]{\lineheight{1.25}\smash{\begin{tabular}[t]{l}$2$\end{tabular}}}}%
    \put(0,0){\includegraphics[width=\unitlength,page=13]{FinalSM.pdf}}%
  \end{picture}%
\endgroup%